\documentclass[ times %,5p%
        ,3p,
        sort&compress ]{elsarticleTC}

\usepackage{graphicx}
\usepackage{amssymb, amsmath}
\usepackage{fancyhdr}
\usepackage[mathscr]{eucal}

\begin{document}

\begin{frontmatter} % elsarticle
%%%
%%% Title page
%%%
\title{Asymmetric contrarians in opinion dynamics}

\author[sg]{Serge Galam}
\ead{serge.galam@sciencespo.fr} % underscore _ requres \
\author[tc]{Taksu Cheon}
\ead{taksu.cheon@kochi-tech.ac.jp} % underscore _ requres \

\address[sg]{
CEVIPOF - Centre for Political Research, Sciences Po and CNRS,
98 rue de l'Universit\'e, 75007 Paris, France} 
\address[tc]{
Laboratory of Physics, Kochi University of Technology, Tosa Yamada, 
Kochi 782-8502, Japan} 

\date{\today}
%\date{June 11, 2019}

\begin{abstract}
Asymmetry in contrarian behavior is investigated within the Galam model of opinion dynamics using update groups of size 3 with two competing opinions A and B. Denoting $x$ and $y$ the respective proportions of A and B contrarians, four schemes of implementations are studied. First scheme activates contrarians after each series of updates with probabilities $x$ and $y$ for agents holding respectively opinion A and B. Second scheme activates contrarians within the update groups only against global majority with probability $x$ when A is majority and $y$ when B is majority. Third scheme considers in-group contrarians acting prior to the local majority update against both local majority and minority opinions. Last scheme activates  in-group contrarians prior to the local majority update but only against the local majority. The main result is the loss of the fifty-fifty attractor produced by symmetric contrarians. Producing a bit less contrarians on its own side than the other side becomes the key to win a public debate, which in turn can guarantee an election victory. The associated phase diagram of opinion dynamics is found to exhibit a rich variety of counterintuitive results.
\end{abstract}

%\pacs{05.10.-a, 87.23.Ge, 89.75.Fb}
%\keywords{opinion dynamics, local and global majority, tipping points, sociophysics.}

\end{frontmatter} % elsarticle
%%%
\section{Introduction}

The modeling of opinion dynamics is one of the major topic of sociophysics studies \cite{celia,bello,carbone,weron,marco,lanchier,iglesias,mobilla,fasano,andre,diep}. Most discrete models consider the dynamics of two competing choices among identical agents, who all follow the same update rule while involved in a discussion with others. For any given initial distribution of choices among the agents, applying repeatedly the update leads to a homogenization of opinions with one of the two competing opinion getting shared by more and more agents. After some number of updates, all agents eventually adopt the same opinion. 

With respect to entropy, the initial random distribution of opinions yields a non zero value of entropy with some associated disorder. However, repeated updates decrease the disorder gradually to end up with a totally ordered state characterized by a zero entropy.

In physical systems, above situation corresponds to a kind of spin system with an initial distribution of up and down spins connected by interactions at zero temperature. Minimizing the energy leads then to a fully ordered ground state with all spins either up or down characterized by a zero entropy. A non-zero temperature increases the disorder with a non zero entropy making the free-energy to be minimized instead of the internal energy.

While the temperature is an essential physical quantity, up to now, it has not been possible to identify a social equivalent. In parallel, the concept of contrarian agents, introduced in the modeling of opinion dynamics within the Galam model \cite{voting, chopa,mino,refe,hetero}, does create a disorder in opinion dynamics, which share features with temperature effects \cite{contra,fifty}. In particular, contrarian agents produces a non zero entropy making attractors to stabilize mixtures of opinions with fixed proportions of them \cite{tosca1,tosca2}. The concept of contrarian behavior has  generated a great deal of works among physicists \cite{bolek,cheon,weron-t,koree,gerard,nuno,bagnoli,zanette} within the active field of scoiophysics \cite{book,santo,review,bikas, testi}.

The contrarian feature is defined as an additional feature to ``normal" behavior of agents following local majority rules for updating their respective opinions once they are discussing in small groups. Contrary to the normal agent, the contrarian behaves in two successive steps. First, it acts as a normal agent by expressing its choice. Then, once a majority has crystallized within the discussing group, while all normal agents adopt it, the contrarian chooses adopting the opposite one. It may be preserving its initial opinions. 

The contrarian shift is independent of either choice being majority. It is also worth to notice that activation of the contrarian behavior is probabilistic. Given an agent  in an updating group, it behaves normally with a probability $(1-c)$ and turns contrarian with probability $c$. Accordingly the same agent may behave normally during an update  and be a contrarian during another update. However at every sequence of simultaneous updates, a proportion $c$ of all agents are contrarians while the complementary proportion $(1-c)$ behaves normally.

In this paper, we extend the concept of contrarian by investigating the effects of making contrarian behavior asymmetric. Instead of one proportion $c$ of contrarians, we considers two different proportions $x$ and $y$ to denote the respective proportions of A and B contrarians \cite{frans-2}. 

This hypothesis implies that the contrarian feature is now related to the choice itself rather than to the agent independently of its current holding choice. It worth to emphasize that asymmetric contrarian behavior departs from physics where asymmetric temperature does not exist. At least, it has not been defined yet.

Our main result is that any non zero value of $\mid x-y \mid$ destroys the hung outcome with the attractor being shifted apart from the fifty-fifty simultaneously restoring a winner over a loser. The corresponding wining strategy is now to have a bit less contrarians on its side than on the other side.

In addition, contrarian asymmetry splits the contrarian activation between two different schemes. Implementing the contrarian shift within the update groups or after the updates before reshuffling yields different outcomes, contrary to the symmetric case where both outcomes are identical. 

The rest of the paper is organized as follows. Next Section review the effect of symmetric contrarians using the Galam model of opinion dynamics. Section 3 introduces asymmetric contrarians with 4 different schemes to activate the contrarian behavior. These schemes are respectively (i) post-update local asymmetric contrarians;
(ii) post-update global asymmetric contrarians; (iii) in-group asymmetric contrarians with majority/minority contrarians; (iv) in-group asymmetric contrarians with in-group majority contrarians. The four different updates equations are discussed in Section 4 while last Section concludes the paper.

%%%%%%%%%%%%%%%%%%%%%%%%%%%%%%%
\section{Symmetric contrarians}

 \subsection{The core model}
 
The contrarian effect has been introduced using the core Galam model of opinion dynamics.  Two opinions A and B are competing within a population of agents denoted floaters. A floater does hold an opinion it has chosen. When a public debate is launched, a floater tries to convince other agents to adopt its own choice. However a floater agent does listen to the arguments supporting the other choice and may be convinced to shift its initial choice. The complicated and complex process of individual making opinion is mimicked by using a local majority rule with every discussing agent having one vote in favor of either one of the two choices.

The underlying dynamic of opinion deploys in threes steps. First, agents are randomly distributed in local groups of size $r$. Then, a local majority rule is applied locally within each group to update individual opinions. It makes each group ending up unanimous in supporting either A or B. In the third step, agents are reshuffled before going back to step 1. Each series of the three steps produces a variation of support for both opinions.  Starting from some initial proportions, $p_0$ and $(1-p_0)$ for respectively A and B, the dynamics yields with $p_0\rightarrow p_1 \rightarrow p_2 \rightarrow \dots$. 

\begin{table}
\centering
\begin{tabular}{|c|c|c|c|}
 \hline
 Configuration & Update &Weight &Probability    \\
 \hline
 A A A $\rightarrow $&  A A A & 1   & $ p^3$ \\
 \hline  \hline
  A A B, A B A, B A A $\rightarrow $&  A A A &1 & $ 3 p^2 (1-p)$ \\
 \hline 
  \hline
 A B B, B A B, B B A $\rightarrow $&  B B B & 0 & $ 3p (1-p)^2 $ \\
 \hline 
 \hline
 B B B  $\rightarrow $&  B B B & 0 & $ (1-p)^3$ \\
 \hline 
 \end{tabular}
\caption{Configuration lists the various distribution of A and B in a group of size $r=3$ for a population with a proportion $p$ in favor of A and $(1-p)$ in favor of B. Update shows the outcome of opinions once majority rule has been applied. Weight gives the corresponding contribution to opinion A. Probability gives the configuration probability.}
\label{ta1}
\end{table}

The updating is iterated till an attractor of the dynamics is reached with $p_0\rightarrow p_1 \rightarrow \dots \rightarrow p_n \rightarrow p_{n+1} \approx p_n$. Defining the local majority function $P(p) $ allows identifying the associated dynamics by solving the fixed points Equation,
\begin{equation}
P(p) = p ,
\label{fp} 
\end{equation}
which yields the  attractors $p_A, p_B$ and tipping point $p_t$ driving the dynamics. Above $p_t$ ($p_0 >p_t$) the support for A increases with $p_0<p_1<p_2<\dots p_n  \approx p_A$, while below $p_t$ ($p_0 < p_t $) the support decreases as $ p_0>p_1>p_2>\dots p_n  \approx p_B$. 

However, in real situations, the vote takes place at some update $l$ before reaching the attractor. The dynamics may also stop by itself with people shifting to discuss another issue after some number $m$ of updates. 

In case of groups of size $r=3$, 8 local configurations are obtained as exhibited inTable (\ref{ta1}). The respective probability of occurrence, the corresponding local majority updates and the resulting weight contributing to opinion A are also given. The  resulting proportion $p_{t+1}$ at $t+1$ is,
\begin{equation}
p_{t+1} \equiv P(p_t)= p_t^3  +  3 p_t^2(1-p_t)=-2p_t^3+3p_t^2 ,
\label{p3} 
\end{equation}
where  $p_{t}$ is the A proportion at $t$. The associated proportion for B is  $(1-p_{t+1})$ with $0 \leq p_{t}\leq \ 1$.

From Eq. (\ref{fp}) two attractors are $p_A=1$ and  $p_B=0$ separated by a tipping point $p_t=\frac{1}{2}$. Therefore the dynamics makes the initial majority to win the debate.

 \subsection{Adding contrarian behavior}

Including the psychological feature of contrarian behavior among floaters can have substantial effects on the dynamics  \cite{contra}. A contrarian departs from a floater by shifting its opinion to oppose the local majority once it has revealed itself. The contrarian shift is independent of the majority opinion being either A or B. A contrarian behavior is activated randomly with some probability $c$, which accounts for the proportion of contrarian behavior within a social group. Contrarian behavior may be constant for some agents and temporary for others. However, at any given time, a proportion $c$ of floaters are contrarian.

The value $c$ is a given fixed external parameter, which is independent of the dynamics and satisfies $0 \leq c \leq \ 1$. Most real cases have $c < \frac{1}{2}$ with $c > \frac{1}{2}$ producing a systematically collective flips.

For size $r=3$, incorporating contrarian shifts once the local majority has been implemented, results in applying contrarian shifts on  unanimous configurations  A A A and B B B. Since contrarian shift is probabilistic,  0, 1, 2, 3 contrarians may be present in configurations A A A and B B B. Table (\ref{ta2}) shows the various cases with the new configurations, the weight contributions to opinion A and the respective probabilities of occurrence, which are $ (1-c)^3, 3 c (1-c)^2, 3 c^2 (1-c), c^3$.

\begin{table}
\centering
\begin{tabular}{|c|c|c|c|c|}
 \hline
 Contrarians &Configuration &Post-update &Weight &Probability    \\
 \hline
0 & A A A $\rightarrow $&  A A A & 1   & $ (1-c)^3$ \\
 \hline  \hline
 1 &  A$_c$ A A ($\times 3$) $\rightarrow $&  B A A & $ \frac{2}{3}$  & $3 c (1-c)^2$ \\
 \hline 
  \hline
 2 & A$_c$ A$_c$ A  ($\times 3$)  $\rightarrow $&  B B A  &$ \frac{1}{3}$   & $3 c^2 (1-c)$ \\
 \hline 
 \hline
 3 & A$_c$ A$_c$ A$_c$  $\rightarrow $&  B B B & 0 & $c^3$ \\
 \hline 
 \hline
 \hline
0 & B B B $\rightarrow $&  B B B & 0   & $ (1-c)^3$ \\
 \hline  \hline
 1 &  B$_c$ B B ($\times 3$) $\rightarrow $&  A B B & $ \frac{1}{3}$  & $3 c (1-c)^2$ \\
 \hline 
 \hline
 2 & B$_c$ B$_c$ B  ($\times 3$)  $\rightarrow $&  A A B  &$ \frac{2}{3}$   & $3 c^2 (1-c)$ \\
 \hline 
 \hline
 3 & B$_c$ B$_c$ B$_c$  $\rightarrow $&  A A A & 1 & $c^3$ \\ \hline 
\end{tabular}
\caption{Distribution of 0, 1, 2, 3 contrarians in configuration A A A once the update has been completed. The associate weight contribution to A and the probability of occurrence are given.}
\label{ta2}
\end{table}

From Table (\ref{ta2}) Equation (\ref{p3}) becomes,
\begin{equation}
P_c(p)= C_{AAA} P(p) + C_{BBB} [1-P(p)],
\label{p3c} 
\end{equation}
with
\begin{eqnarray}
C_{AAA}& = & (1-c)^3+\frac{2}{3}3c(1-c)^2+\frac{1}{3}3c^2(1-c) \nonumber\\
& = & (1-c) ,
\label{Caaa} 
\end{eqnarray}
and 
\begin{eqnarray}
C_{BBB}& = & \frac{1}{3}3c(1-c)^2+\frac{2}{3}3c^2(1-c) +c^3 \nonumber\\
& = & c .
\label{Cbbb} 
\end{eqnarray}

Using Equations (\ref{Caaa}, \ref{Cbbb}), the update Equation (\ref{p3c}) writes as,
\begin{equation}
P_c(p)= (1-c) \Big[ p^3  +  3 p^2(1-p)\Big]+ c\Big[3p(1-p)^2+(1-p)^3\Big] ,
\label{pp3c} 
\end{equation}
which shows that activation of contrarians could be also applied identically to individual agents out of the update groups. Equation (\ref{pp3c}) simplifies to,
\begin{equation}
P_c(p)= (1-2c)\Big[-2p_t^3+3p_t^2\Big] +c ,
\label{pp3cc}
\end{equation}
which yields two attractors $p_A, p_B$ and a tipping point $p_t$ defined as,
\begin{eqnarray}
&p_A=& \frac{(1-2c) +\sqrt {1-8c+12c^2 }}  {2(1-2c)}, \nonumber\\
&p_B=& \frac{(1-2c) -\sqrt {1-8c+12c^2 }}  {2(1-2c)} ,\nonumber\\
& p_t= & \frac{1}{2}  ,
\label{p6} 
\end{eqnarray}
provided $0\leq c \leq  \frac{1}{6} \approx 0.1667$ and $ \frac{1}{2}  \leq c \leq1$ to ensure the existence of the square root.

\begin{figure}
\centering
\includegraphics[width=.41\textwidth]{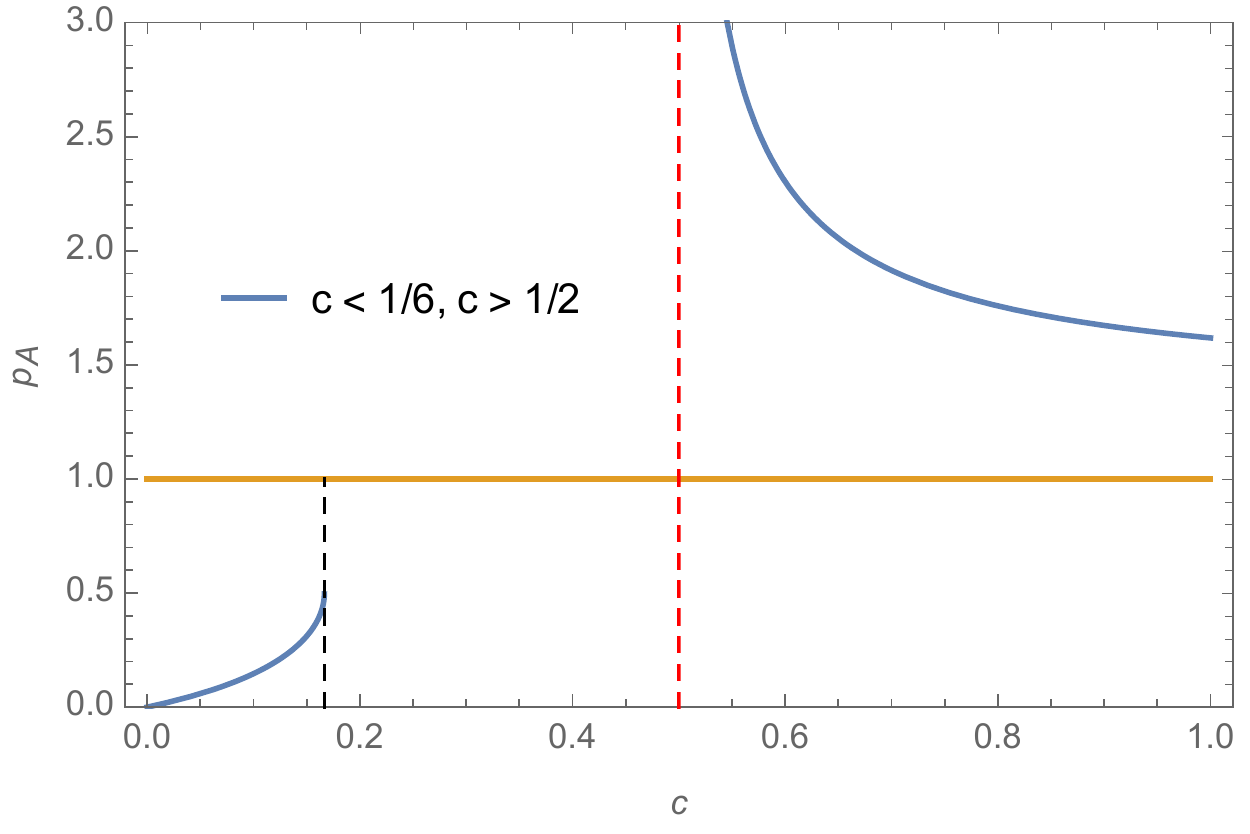}\ \ \ \  
\includegraphics[width=.41\textwidth]{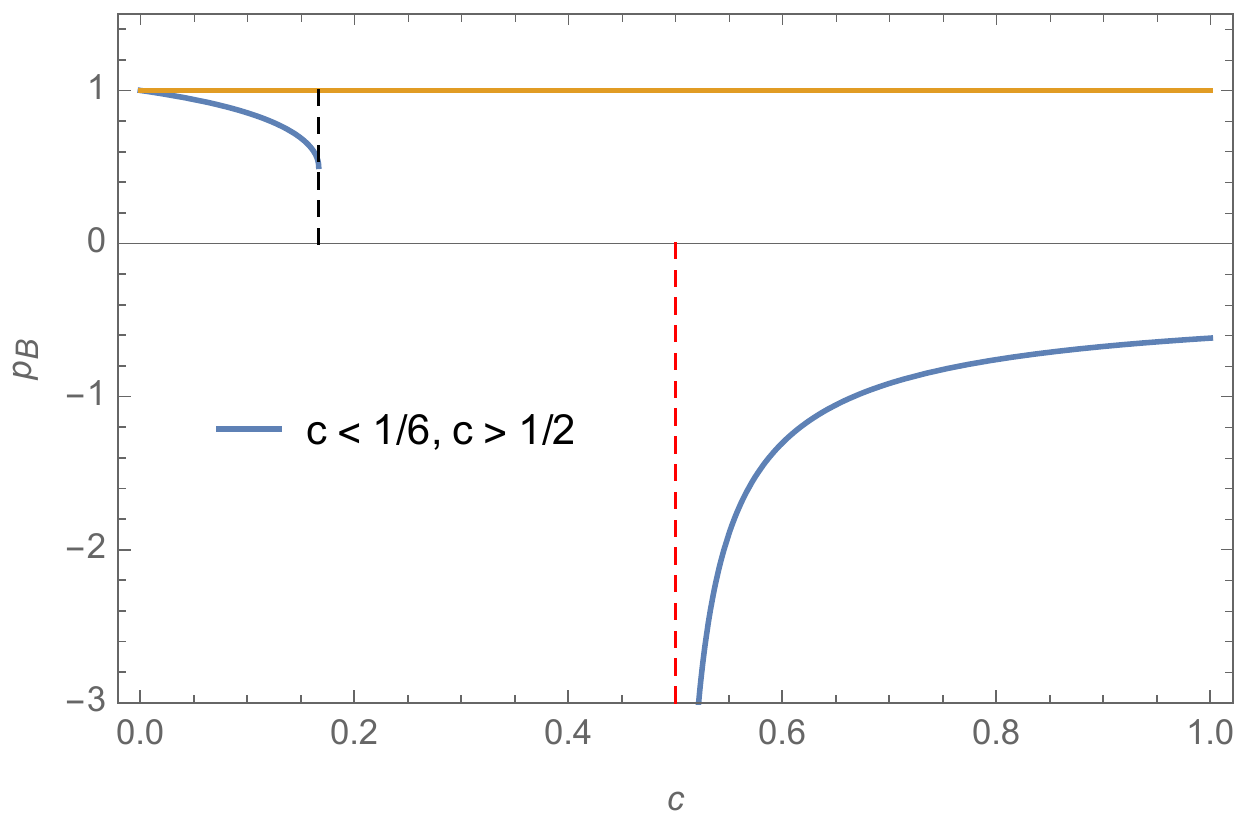}
\caption{Variations of $p_A$ (left) and $p_B$ (right) as a function of $c$ are shown. The two attractors exist only in the range $0\leq c \leq 1$.}
\label{pABc}
\end{figure}

\begin{figure}
\centering
\includegraphics[width=.34\textwidth]{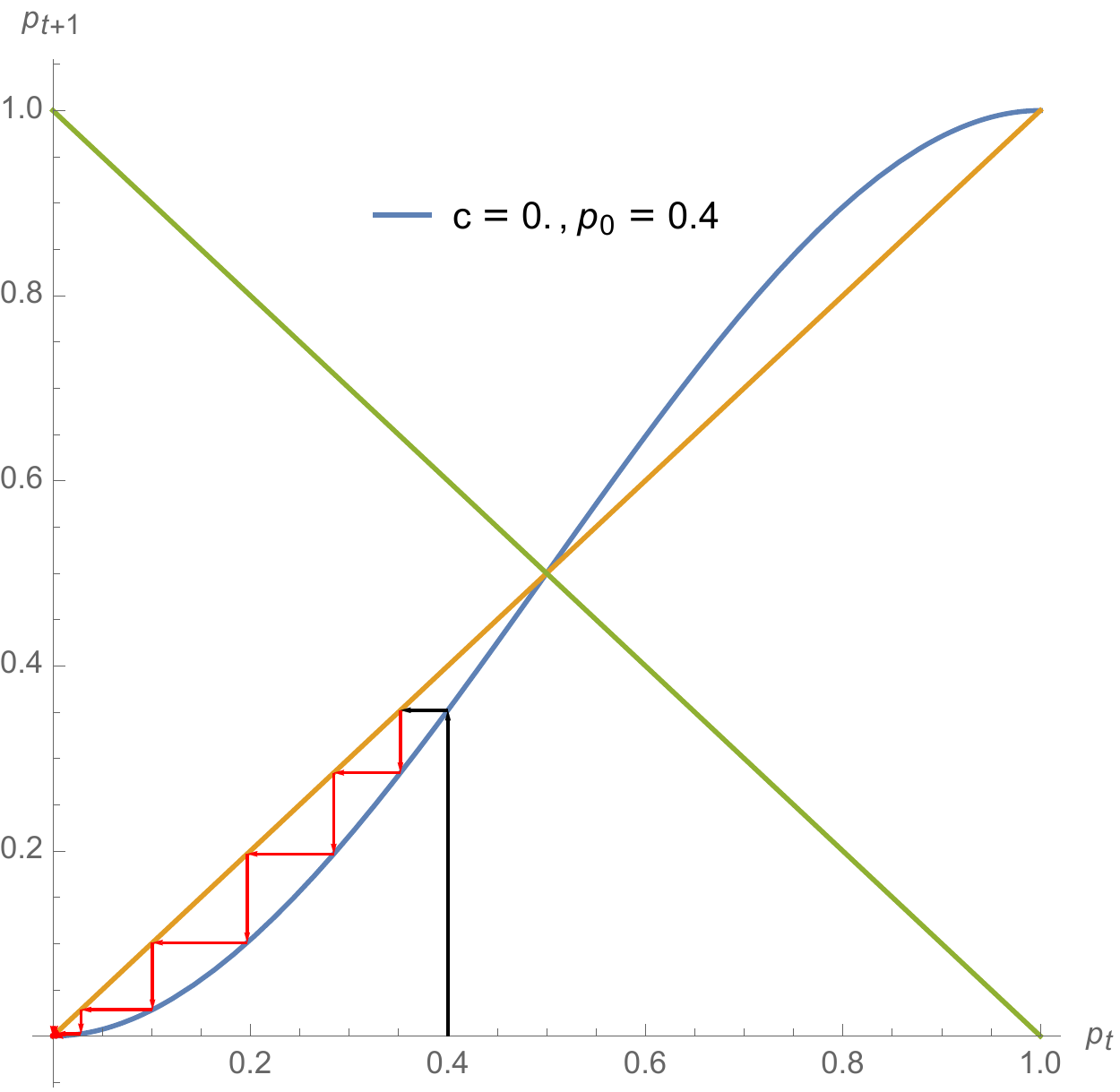}\ \ \ \ 
\includegraphics[width=.34\textwidth]{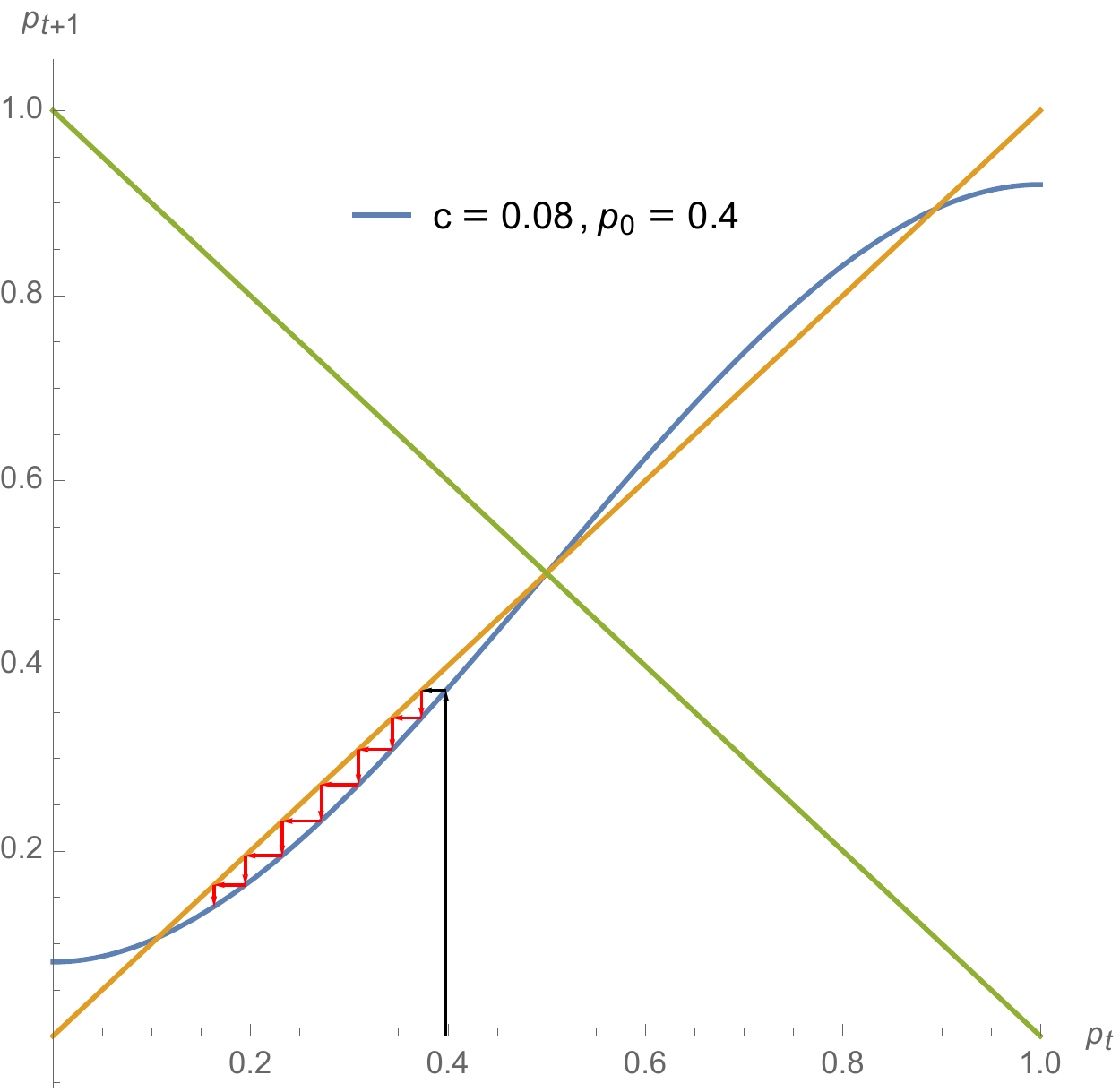}
\caption{Using Eq.(\ref{pp3cc}) the expression $p_{t+1}$ is shown as a function of $p_t$ for $c=0$ (left) and $c=0.08$ (right). The arrows show the dynamics from $p_0=0.4$ for several iterations. Associated attractors are $p_A=0, p_B=1, p_t=0.50$ for the first case and  $p_A=0.11, p_B=0.89, p_t=0.50$ for the second case.}
\label{c01}
\end{figure}

\begin{figure}
\centering
\includegraphics[width=.34\textwidth]{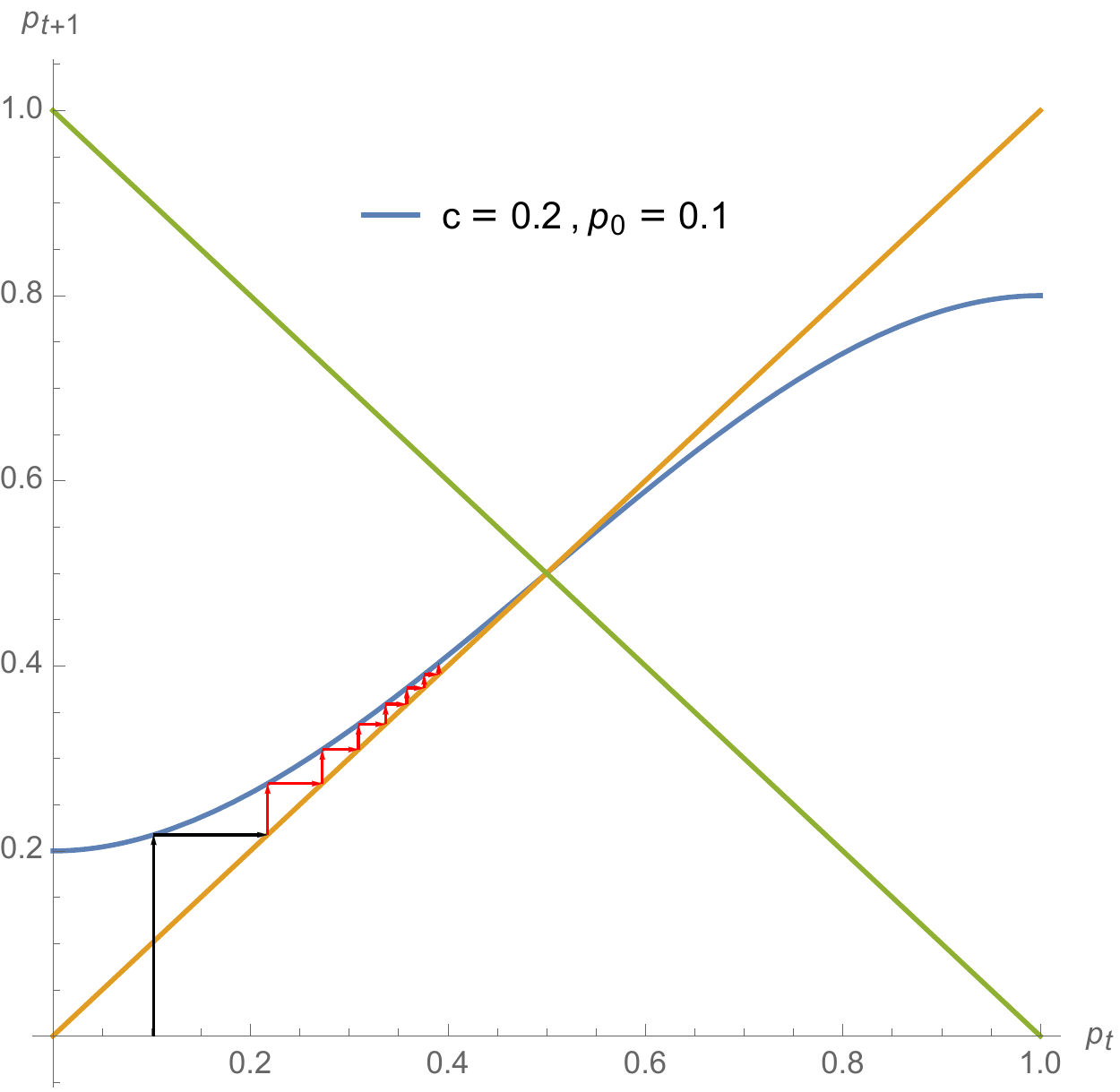}\ \ \ \ 
\includegraphics[width=.34\textwidth]{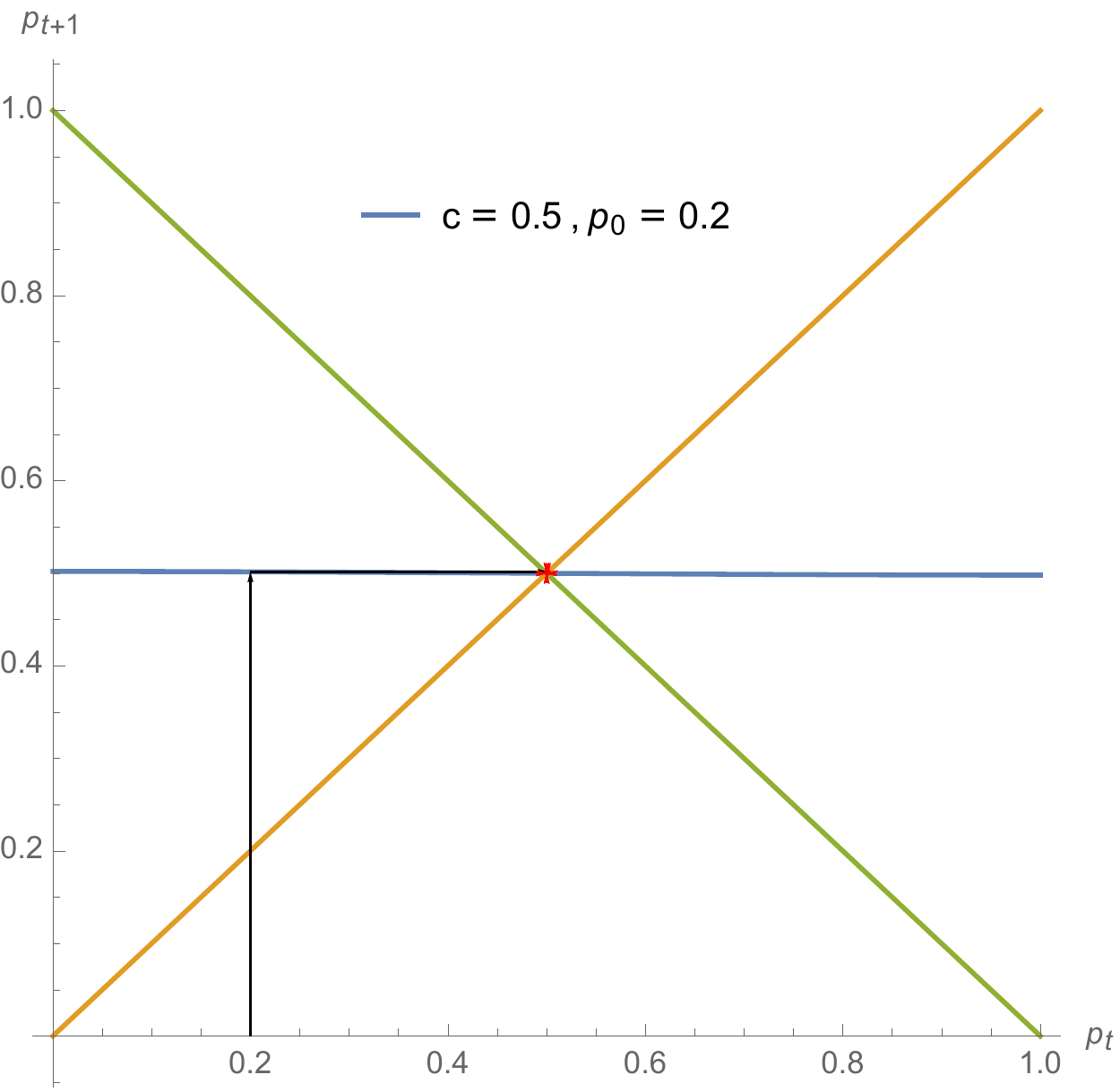}
\caption{Using Eq.(\ref{pp3cc}) the expression $p_{t+1}$ is shown as a function of $p_t$ for $c=0.2$ (left) and $c=0.5$ (right). The arrows show the dynamics from $p_0=0.1$ (left) for several iterations. For the second case, any initial value $p_0$ leads to the single attractor within one iteration. For both cases one single attractor $p_A=p_B=p_t=0.50$ drives the dynamics.}
\label{c23}
\end{figure}

However for $ \frac{1}{2}  < c < 1$ we have $p_A>1$ and $p_B<0$ making them non physical. At c=$\frac{1}{6}$ and $c=\frac{1}{2}$, $p_A=p_B=\frac{1}{2}$. Variations of   $p_A$ and $p_B$ as a function of $c$ are shown in Figure (\ref{pABc}). Indeed, for $c \geq  \frac{1}{6}, $ $p_A$ and $p_B$ merge at the tipping point $p_t = \frac{1}{2}$ turning it to the unique attractor of the dynamics. The two regimes are shown in Figures (\ref{c01}, \ref{c23}) with $c=0.08 \rightarrow p_A=0.11, p_B=0.89, p_t=0.50$ and $c=0.20 \rightarrow p_A=p_B=p_t=0.50$. For the full range $c> \frac{1}{6}$ any initial value $p_0$ ends up at  $\frac{1}{2}$. For $c=\frac{1}{2}$ we have $p_{t+1}= \frac{1}{2}$ for whatever value of $p_t$.

\begin{figure}
\centering
\includegraphics[width=.34\textwidth]{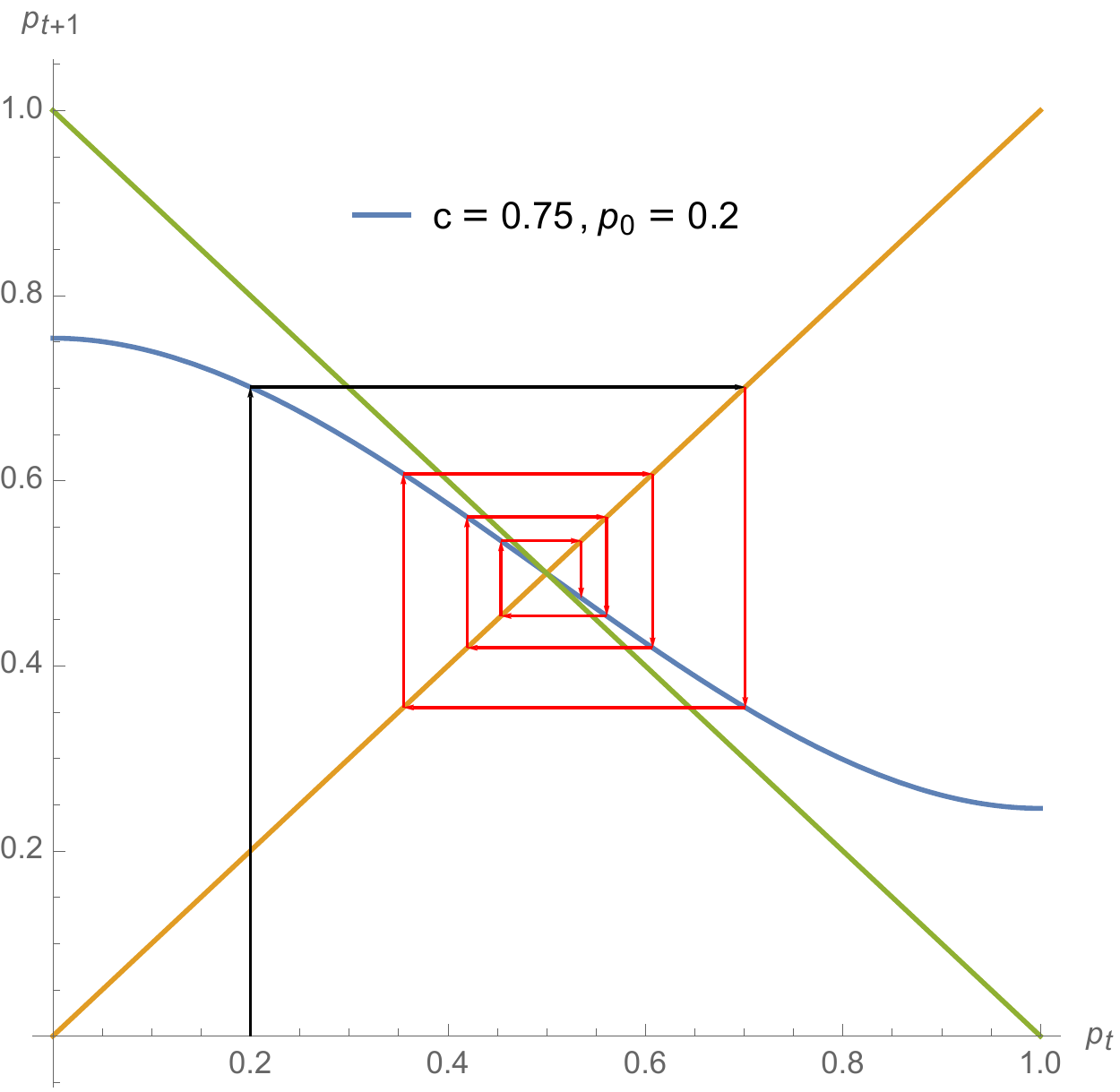}\ \ \ \ 
\includegraphics[width=.34\textwidth]{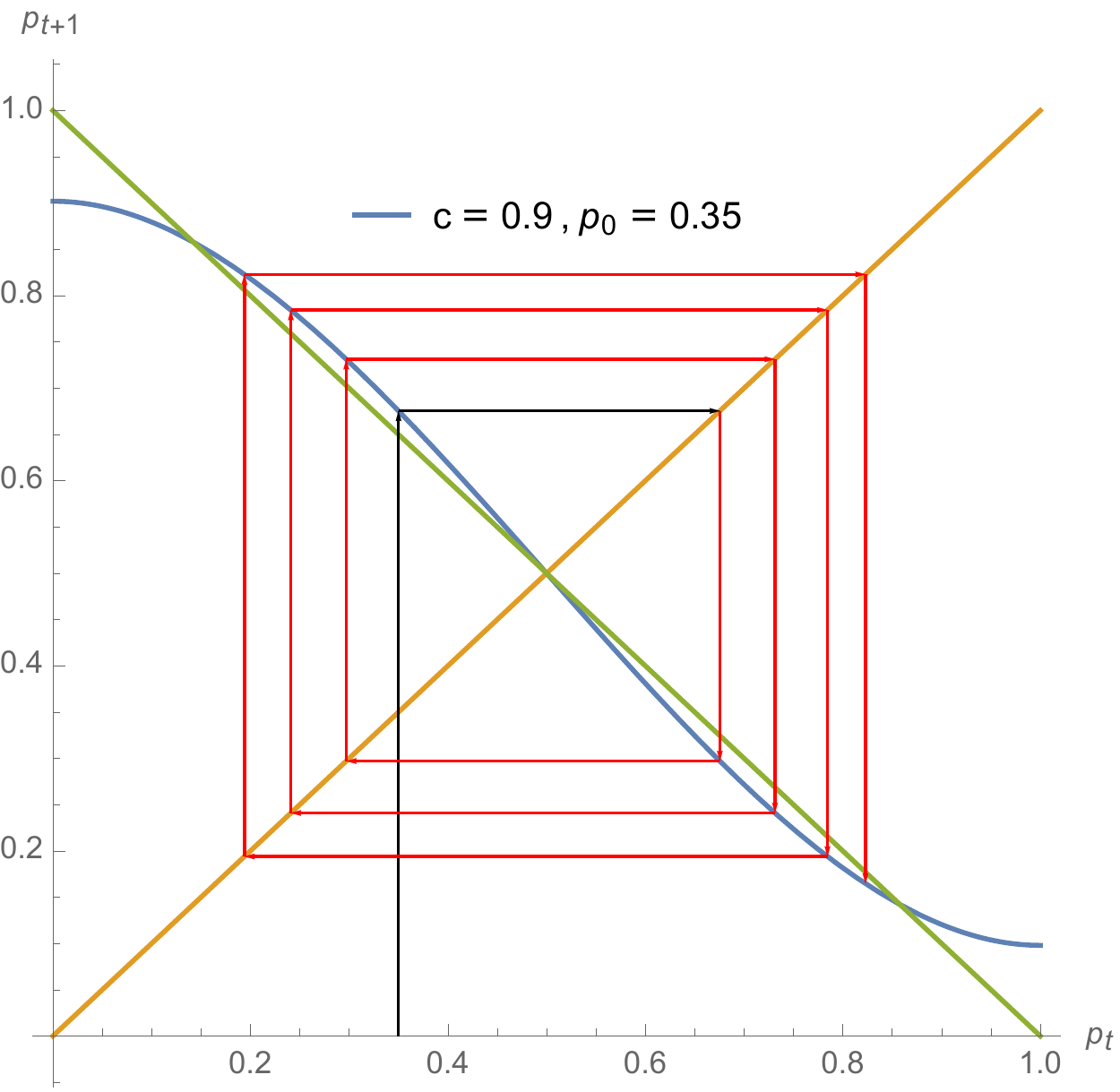}
\caption{As soon as $c>0.5$ the dynamics becomes alternating. Using Eq.(\ref{pp3cc}) the expression $p_{t+1}$ is shown as a function of $p_t$ for $c=0.75$ (left) and $c=0.9$ (right). The arrows show the dynamics from $p_0=0.1$ (left) and $p_0=0.35$ (right) for several iterations. For the first case, any initial value $p_0$ leads to the single attractor $p_A=p_B=p_t=0.50$  but through jumps in respective supports for both opinions. For the second case, the iteration produces jumps around the tipping point $p_t=0.50$ but now alternating towards  $p_A=0.15$ and $p_B=0.85$.}
\label{c45}
\end{figure}

The different regimes of opinion dynamics driven by a non zero proportion of contrarian behavior are exhibited in the series of Figures  (\ref{c01}, \ref{c23}, \ref{c45}).

\section{Making contrarians asymmetric}

Above symmetric contrarian behavior could be connected to the assumption that contrarianism is an individual feature produced by some social environment. It is individuals a psychological feature to oppose the surrounding leading opinion. It is therefore independent of opinions themselves, which justifies making contrarian behavior symmetrical with respect to A and B  \cite{contra}.

However, contrarianism could be also identified as arising from the cultural content in which a given opinion is rooted. Behind an opinion choice about a given issue would stand a global frame of thinking. In this case, contrarianism could become asymmetric, some social frames favoring more than others, an individual inclination to challenge the leading opinion adopting the minority view. Denoting $x$ the probability that an A holding contrarian flips to B and $y$ the probability that a B holding contrarian flips to A, we study the effect of such an asymmetry on the associated dynamics of opinion. 

It is worth to notice that with symmetrical contrarians, the activation of contrarian shift was found identical when applied within the update groups or after the updates during the reshuffling prior to a new update. However, with asymmetric contrarians, while the equivalence is still valid for unanimity configurations A A A and B B B, it does not hold for configurations A A B  and A B B.

Having $x\neq y$ makes simple analytical solving not possible to determine the values of the various attractors and tipping points as when $x=y$. However, it will be seen that the two regimes associated with symmetric contrarians, one with two attractors and a tipping poing and another one with a single attractor, are still prevailing but with the loss of symmetry for the location of attractors and tipping points.

Accordingly, in the following of the paper, to keep the presentation clear, both regimes of the dynamics will be illustrated choosing specific set of values for $x$ and $y$, which exhibit the main shape of the corresponding phase diagram.

\subsection{Post-update local asymmetric contrarians}

When contrarians are activated after the update, Equation (\ref{pp3c}) is modified as,
\begin{equation}
P_{x,y}^1(p)= (1-x) \Big[ p^3  +  3 p^2(1-p)\Big]+ y\Big[3p(1-p)^2+(1-p)^3\Big] ,
\label{p7}
\end{equation}
which shows that $x\neq y$ main effect is twofold removing the hung election scenario (single attractor at fifty/fifty) yielded by Eq. (\ref{p6}) and turning the dynamics unbalanced. The shift of p$_A$  (p$_B$) towards p$_t$ becomes faster than the shift of p$_B$ (p$_A$) when $x>y$ ($x<y$) since now p$_t$ is also moving departing from $\frac{1}{2}$ to lower  (higher) values.

\begin{figure}
\centering
\includegraphics[width=.41\textwidth]{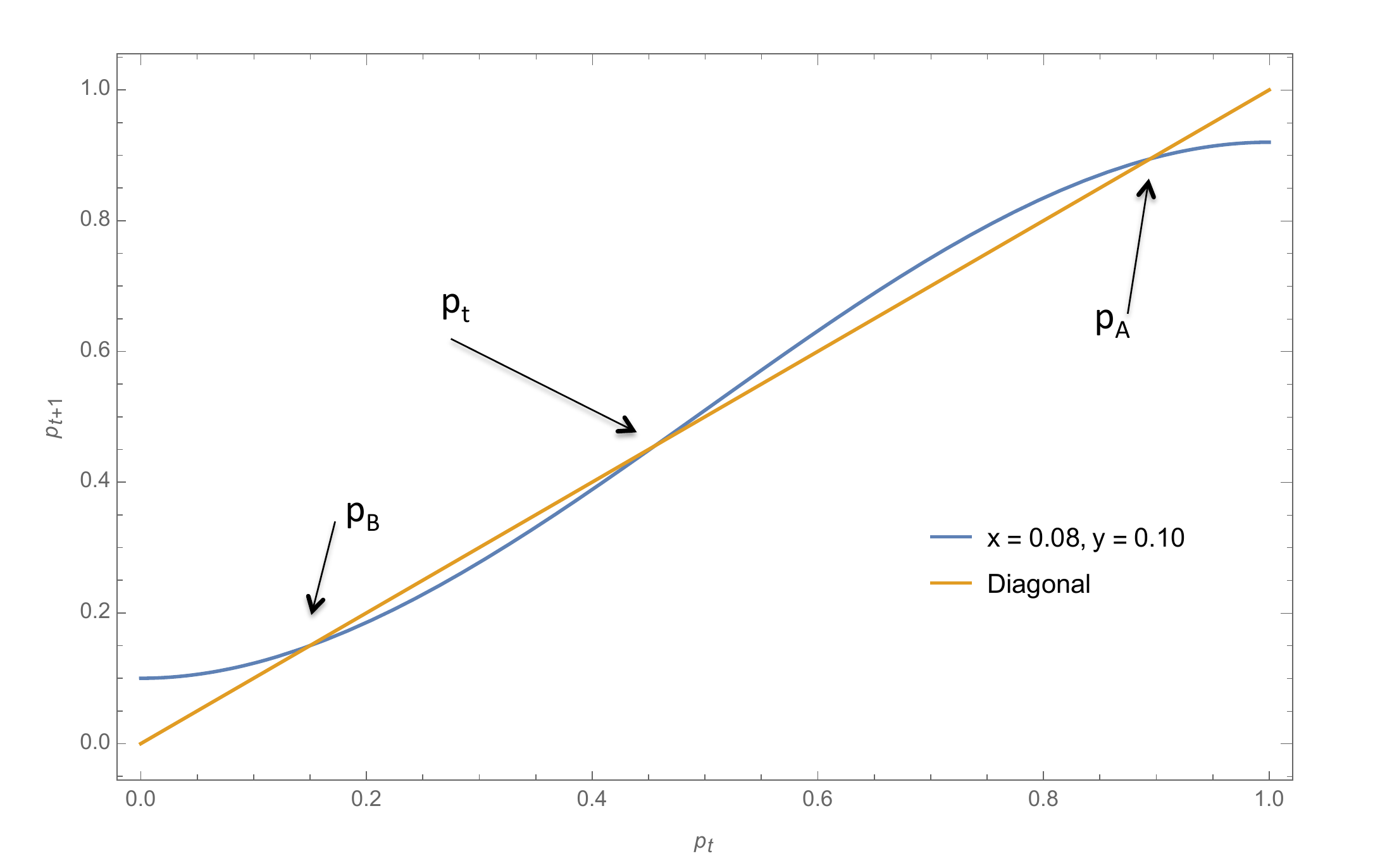}\ \ \ \ 
\includegraphics[width=.41\textwidth]{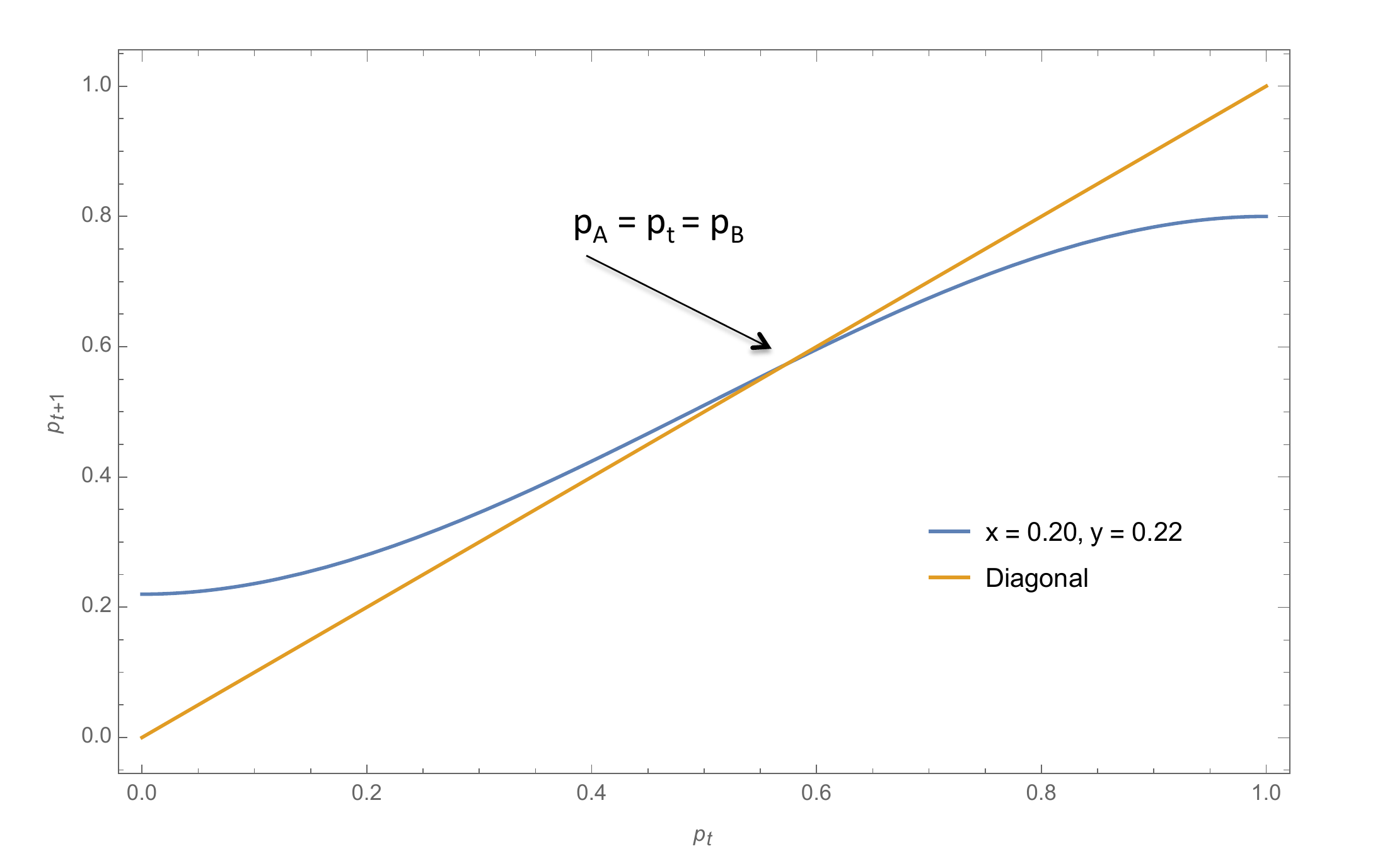}
\caption{Using Eq. (\ref{p7}) the expression $p_{t+1}$ as a function of $p_t$ is shown for $x=0.08, y=0.10$ (upper curve) and $x=0.20, y=0.22$ (lower curve).
The associated fixed points are respectively $p_A=0.15, p_t=0.46, p_B=0.89$ and $p_A=p_t=p_B=0.57$.} 
\label{f3f4}
\end{figure}

For instance in the case $x=0.08$ and  $y=0.10$, Eq. (\ref{p7}) yields $p_A=0.15, p_t=0.46, p_B=0.89$. The coalescence phenomena observed with symmetrical contrarians is still valid but now it occurs at a value different from $\frac{1}{2}$ as seen from the case $x=0.20, y=0.22$ yielding $p_A=p_t=p_B=0.57$. These two cases are exhibited in Figure  (\ref{f3f4}).

\subsection{Post-update global asymmetric contrarians }

A variant of post-update contrarians has been studied in the symmetrical case considering the majority status of an opinion at the global level instead of the local one \cite{contra-global}.  Namely, a contrarian flip from A to B takes place with probability $c$ only when $p>\frac{1}{2}$ and from B to A with the same probability $c$ only when $p<\frac{1}{2}$. In this case, a chaotic behavior is found around an attractor at fifty/fifty. 

Within this global scheme, considering asymmetric contrarians makes  a contrarian flip from A to B to occur with probability $x$ only $p>\frac{1}{2}$. When $p<\frac{1}{2}$ contrarian flips occur only from B to A with probability $y$. The associated update Equation writes as,
\begin{eqnarray}
p>\frac{1}{2} \rightarrow P_{x}^{2a}(p)&=&  (1-x) \Big[ p^3  +   3 p^2(1-p) \Big] , \nonumber\\ 
p<\frac{1}{2} \rightarrow P_{y}^{2b}( (p)&=&   p^3  +  3 p^2(1-p) + y \Big[ (1-p)^3  +   3 p(1-p)^2 \Big],
\label{8}
\end{eqnarray}
which still produces a chaotic behavior but now at a value of $p_t \neq \frac{1}{2}$ instead of $p_t = \frac{1}{2}$. 

\begin{figure}
\centering
\includegraphics[width=.41\textwidth]{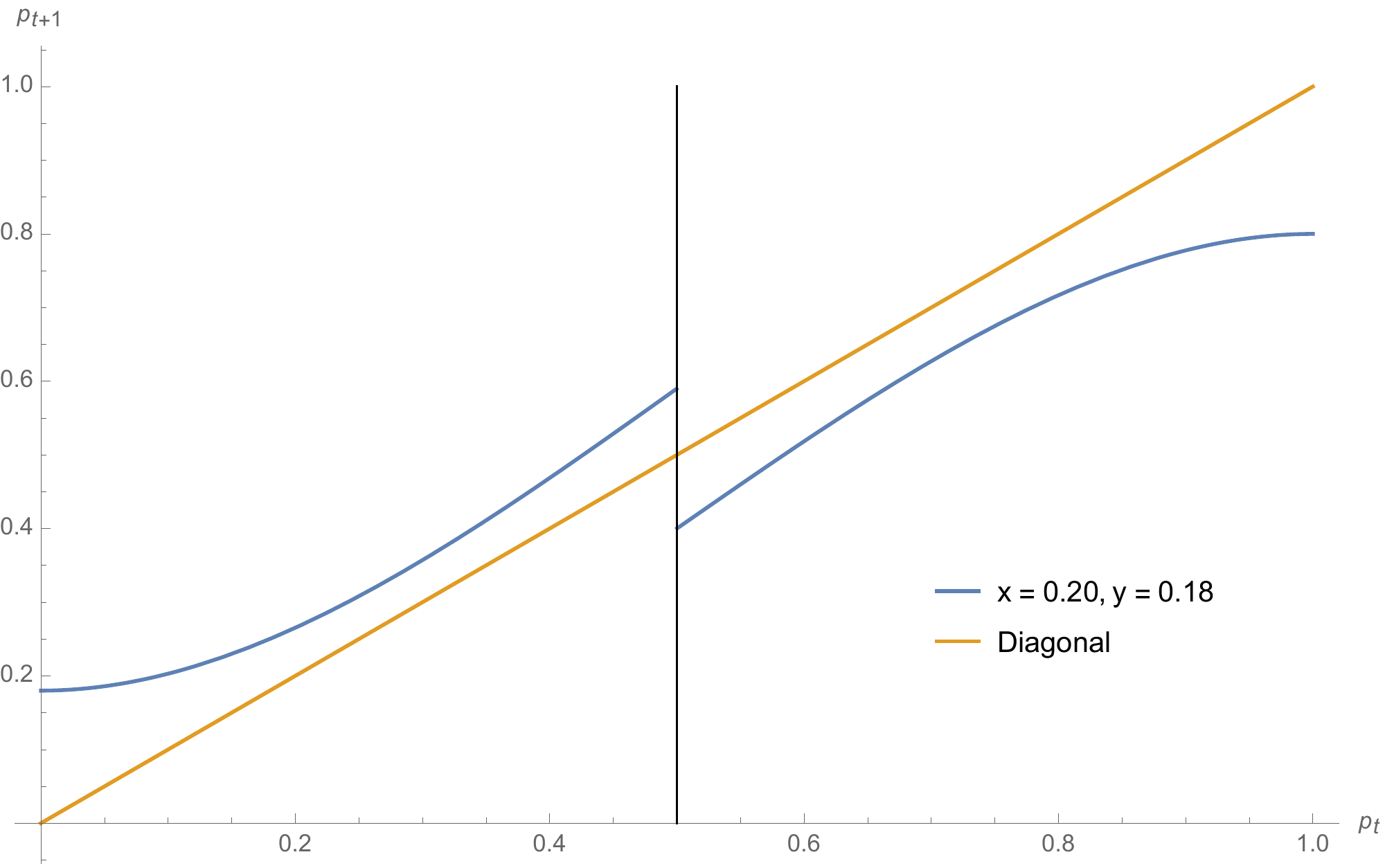}\ \ \ \ 
\includegraphics[width=.41\textwidth]{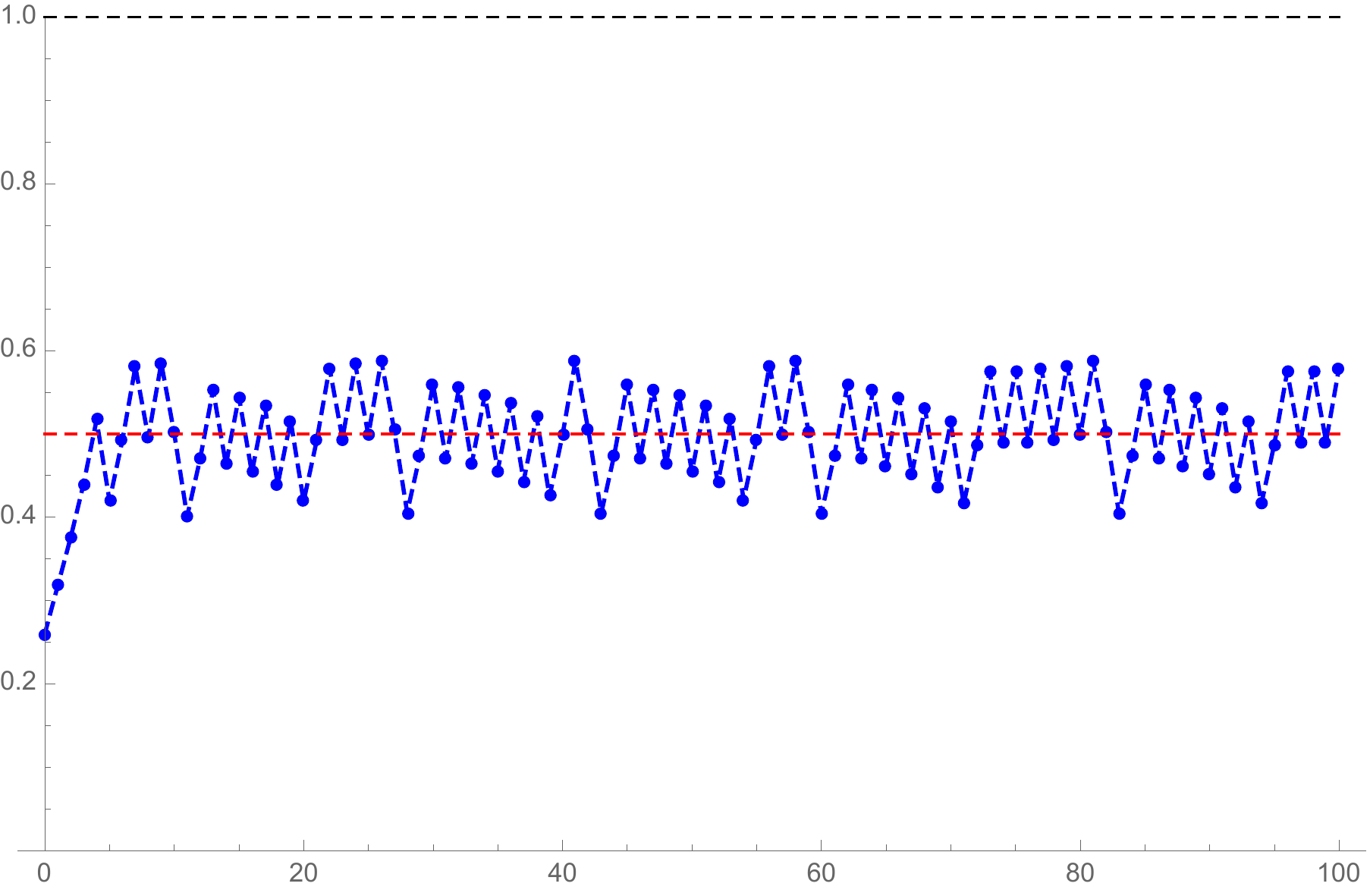}
\caption{Using Eq. (\ref{p8}) the expression $p_{t+1}$ as a function of $p_t$ is shown for $x=0.20, y=0.18$ (upper curve). The associated evolution from $p_0=0.26$ is exhibited (lower curve). The attractor $p_t = \frac{1}{2}$ has split in two limiting points located at $p_t^a = 0.4$ and $p_t^b=0.59$.} 
\label{ca1ca2}
\end{figure}

Figure (\ref{ca1ca2}) illustrates the behavior from Eq. (\ref{p8}) for the case $x=0.20, y=0.18$. The attractor $p_t = \frac{1}{2}$ has split in two limiting points located at $p_t^a = 0.4$ and $p_t^b=0.59$.

\subsection{In-group asymmetric contrarians}

Activation of contrarians inside the update group opens two different possible schemes since A and B agents are not in the same situation as seen with the case A A B$_c$ (A$_c$ B B). First scheme considers that B$_c$ (A$_c$ ) being already minority against A (B), keeps the same opinion B (A). On the other hand, the second scheme considers that B$_c$ (A$_c$ ) being minority, its contrarian feature is not activated and thus the agent follows the local majority turning A (B). However, for configurations A A A and B B B, there is no change with respect to the post-update scheme.

\subsubsection{In-group majority/minority contrarians}
First scheme was investigated in \cite{frans-2} with for instance the update A A B$_c \rightarrow$ A A B. All configuration updates with independent contrarians are listed in Table (\ref{ta3}) for groups of size 3 for the configuration A A B with 0, 1, 2, 3 contrarians.

\begin{table}
\centering
\begin{tabular}{|c|c|c|c|c|}
 \hline
 Contrarians &Configuration & In-group majority/minority  &Weight &Proba    \\
 \hline
 \hline
0 & A A B $\rightarrow $&  A A A & 1   & $ (1-x)^2(1-y)$ \\
 \hline  \hline
 1 &  A$_c$ A B $\rightarrow $&  B A A & $ \frac{2}{3}$  & $ 2 x (1-x)(1-y)$ \\
 \hline 
1 &  A A B$_c$  $\rightarrow $&  A A B &$\frac{2}{3}$ & $ (1-x)^2 y$ \\
 \hline 
 \hline
 2 & A$_c$ A$_c$  B $\rightarrow $&  B B A & $\frac{1}{3}$  & $ x^2(1-y)$ \\
 \hline 
2 & A$_c$  A B$_c$  $\rightarrow $&  B A B & $  \frac{1}{3}$  & $ 2 x (1-x) y$ \\
 \hline 
 \hline
 3 & A$_c$ A$_c$ B$_c$ $\rightarrow $&  B B B & 0  & $ x^2 y$ \\
 \hline
 \end{tabular}
\caption{Distribution of 0, 1, 2, 3 contrarians within a configuration A A B. Associated updates are shown with the weight contribution to A and the probability of occurrence.}
\label{ta3}
\end{table}

The update Equation (\ref{p3}) writes now,
\begin{eqnarray}
P_{x,y}^3(p)&=&  C_{AAA}(x) p^3  + C_{AAB}(x,y) 3 p^2(1-p)  \nonumber\\
& + &   C_{BBB}(y) (1-p)^3 +C_{ABB}(x,y) 3 p(1-p)^2 ,
\label{p8} 
\end{eqnarray}
with from Table  (\ref{ta2}),
\begin{eqnarray}
C_{AAA}(x) & = & (1-x)^3  + 2 x (1-x)^2  + x^2 (1-x) +  0 x^3 \nonumber\\
& = & (1-x) ,
\label{p9}
\end{eqnarray} 
and from Table  (\ref{ta3}),
\begin{eqnarray}
C_{AAB}(x,y) & = &(1-x)^2 (1-y)  \nonumber\\
& + &  \frac{4}{3} x (1-x) (1-y)  +  \frac{2}{3} (1-x)^2 y  \nonumber\\
& +  &\frac{1}{3} x^2 (1-y) + \frac{2}{3}x (1-x) y \nonumber\\
& = & 1 - \frac{2}{3} x - \frac{1}{3} y   \ ,
\label{p10}
\end{eqnarray}
By symmetry we get 
\begin{equation}
C_{BBB}(y) =y
\label{p11} 
\end{equation}
and
\begin{equation}
C_{BBA}(x,y) =  \frac{2}{3} y + \frac{1}{3} x ,
\label{p12} 
\end{equation}
to yield,
\begin{eqnarray}
P_{x,y}^3(p) & = & (1-x) p^3  +  y (1-p)^3 \nonumber\\
& + & \Big[ 1- \frac{2}{3} x - \frac{1}{3} y) \Big] 3 p^2(1-p)   \nonumber\\
& + & \Big[ \frac{2}{3} y + \frac{1}{3} x\Big] 3 p(1-p)^2 \ ,
\label{p13}
\end{eqnarray}
which writes, 
\begin{equation}
P_{x,x}^3(p)=  (1-x) \Big[ p^3  +  3 p^2(1-p) \Big] + x \Big[ (1-p)^3  +  3 p(1-p)^2 \Big],
\label{p14} 
\end{equation}
when $x=y$ thus recovering Eq. (\ref{pp3c}) with $x=c$.
 
\begin{figure}
\centering
\includegraphics[width=.41\textwidth]{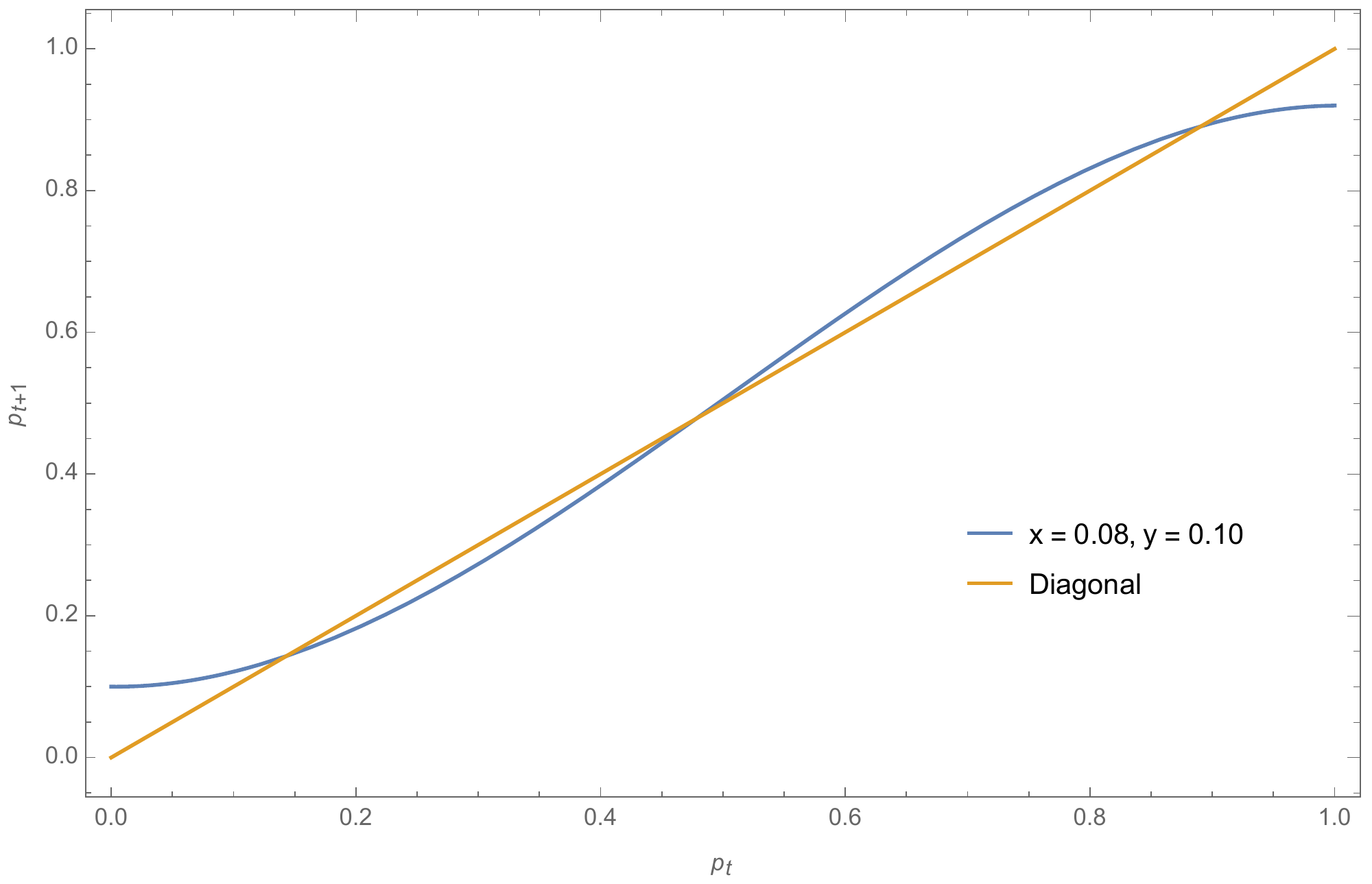}\ \ \ \ 
\includegraphics[width=.41\textwidth]{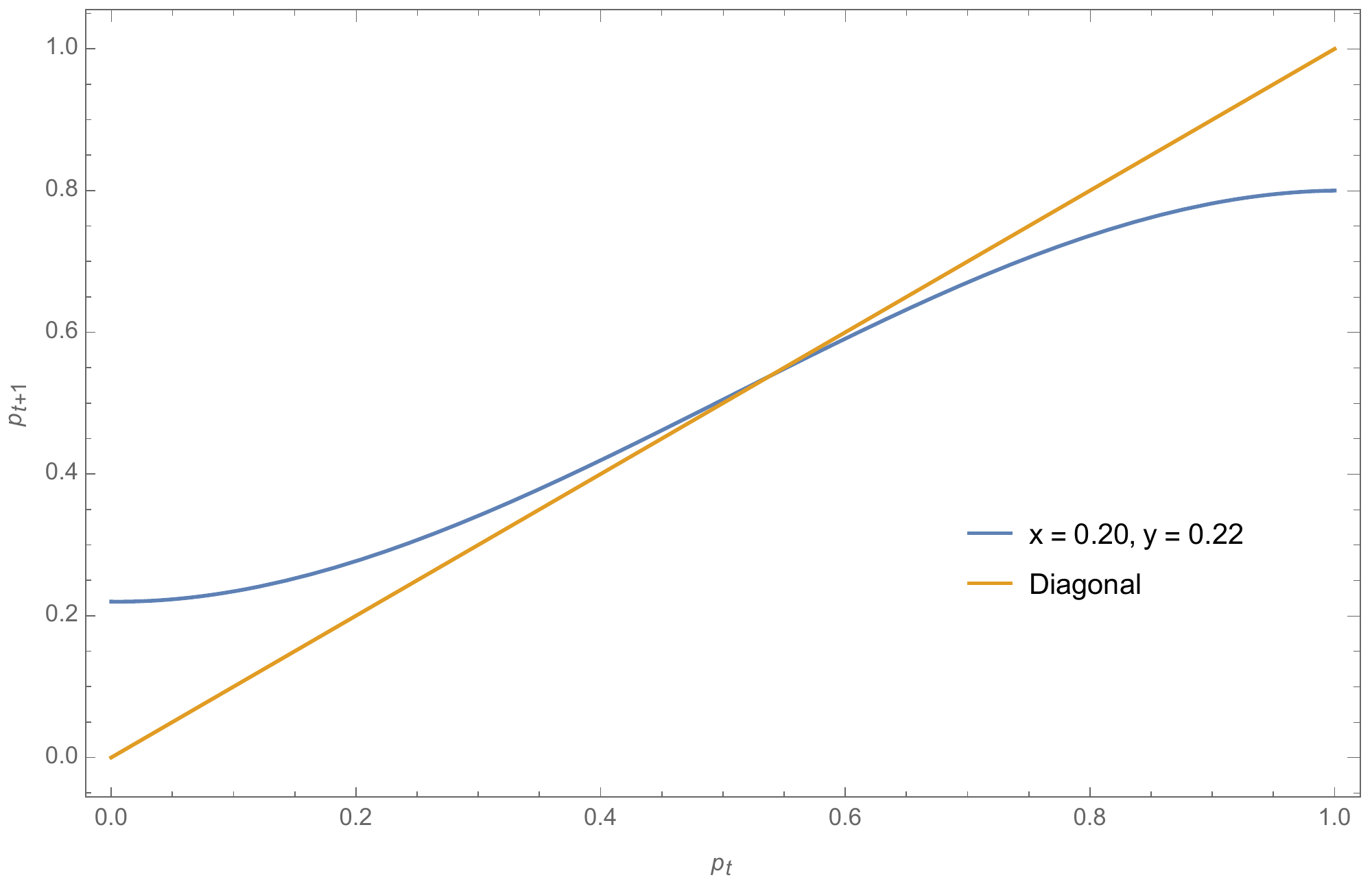}
\caption{Using Eq. (\ref{p13}) the expression $p_{t+1}$ as a function of $p_t$ is shown for $x=0.08, y=0.10$ (upper curve) and $x=0.20, y=0.22$ (lower curve).
The associated fixed points are respectively $p_A=0.14, p_t=0.48, p_B=0.89$ and $p_A=p_t=p_B=0.54$.} 
\label{f5f6}
\end{figure}

For $x=0.08$ and  $y=0.10$, Eq. (\ref{p13}) yields $p_A=0.14, p_t=0.48, p_B=0.89$. The coalescence phenomena observed with symmetrical contrarians is still valid but now it occurs at a value different from $\frac{1}{2}$ as seen from the case $x=0.20, y=0.22$ yielding $p_A=p_t=p_B=0.54$. These two cases are exhibited in Figure  (\ref{f5f6}).

\subsubsection{In-group majority contrarians}

While above, contrarians are always activated within the update group, we could consider that contrarianism is activated at a contrarian agent only when this agent finds itself in an initial local majority in a group. In this case, a contrarian B$_c$ (A$_c$) in a group A A B$_c$ (B B A$_c$) keeps dormant and acts as a regular floater following the local majority rule. According to this scheme  A A B$_c$ (B B A$_c$) $\rightarrow$ A A A (B B B) instead of A A B (B B A) above. Table  (\ref{ta3}) is thus modified as shown in Table  (\ref{ta4}). Accordingly, Eq. (\ref{p10})  becomes,
\begin{eqnarray}
C_{AAB}(x,y) & = &(1-x)^2 (1-y)  \nonumber\\
& + &  \frac{4}{3} x (1-x) (1-y)  +  (1-x)^2 y  \nonumber\\
& +  &\frac{1}{3} x^2 (1-y) + \frac{4}{3}x (1-x) y + \frac{1}{3} x^2y\nonumber\\
& = & 1 - \frac{2}{3} x  \ ,
\label{p15}
\end{eqnarray}
and by symmetry Eq. (\ref{p12})  becomes,
\begin{equation}
C_{BBA}(x,y) =  \frac{2}{3} y  ,
\label{p16} 
\end{equation}
to yield,
\begin{eqnarray}
P_{x,y}^4(p) & = & (1-x) p^3  +  y (1-p)^3 \nonumber\\
& + & \Big[ 1- \frac{2}{3} x) \Big] 3 p^2(1-p)   \nonumber\\
& + &  \Big[  \frac{2}{3} y  \Big]  3 p(1-p)^2 \ .
\label{p15}
\end{eqnarray}

\begin{table}
\centering
\begin{tabular}{|c|c|c|c|c|}
 \hline
 Contrarians &Configuration & In-group majority/minority  &Weight &Proba    \\
 \hline
 \hline
0 & A A B $\rightarrow $&  A A A & 1   & $ (1-x)^2(1-y)$ \\
 \hline  \hline
 1 &  A$_c$ A B $\rightarrow $&  B A A & $ \frac{2}{3}$  & $ 2 x (1-x)(1-y)$ \\
 \hline 
1 &  A A B$_c$  $\rightarrow $&  A A A &$ 1 $ & $ (1-x)^2 y$ \\
 \hline 
 \hline
 2 & A$_c$ A$_c$  B $\rightarrow $&  B B A & $\frac{1}{3}$  & $ x^2(1-y)$ \\
 \hline 
2 & A$_c$  A B$_c$  $\rightarrow $&  B A A & $  \frac{2}{3}$  & $ 2 x (1-x) y$ \\
 \hline 
 \hline
 3 & A$_c$ A$_c$ B$_c$ $\rightarrow $&  B B A & $\frac{1}{3}$  & $ x^2 y$ \\
 \hline
 \end{tabular}
\caption{Distribution of 0, 1, 2 contrarians within a configuration A A B. Associated updates are shown with the weight contribution to A and the probability of occurrence.}
\label{ta4}
\end{table}

\begin{figure}
\centering
\includegraphics[width=.41\textwidth]{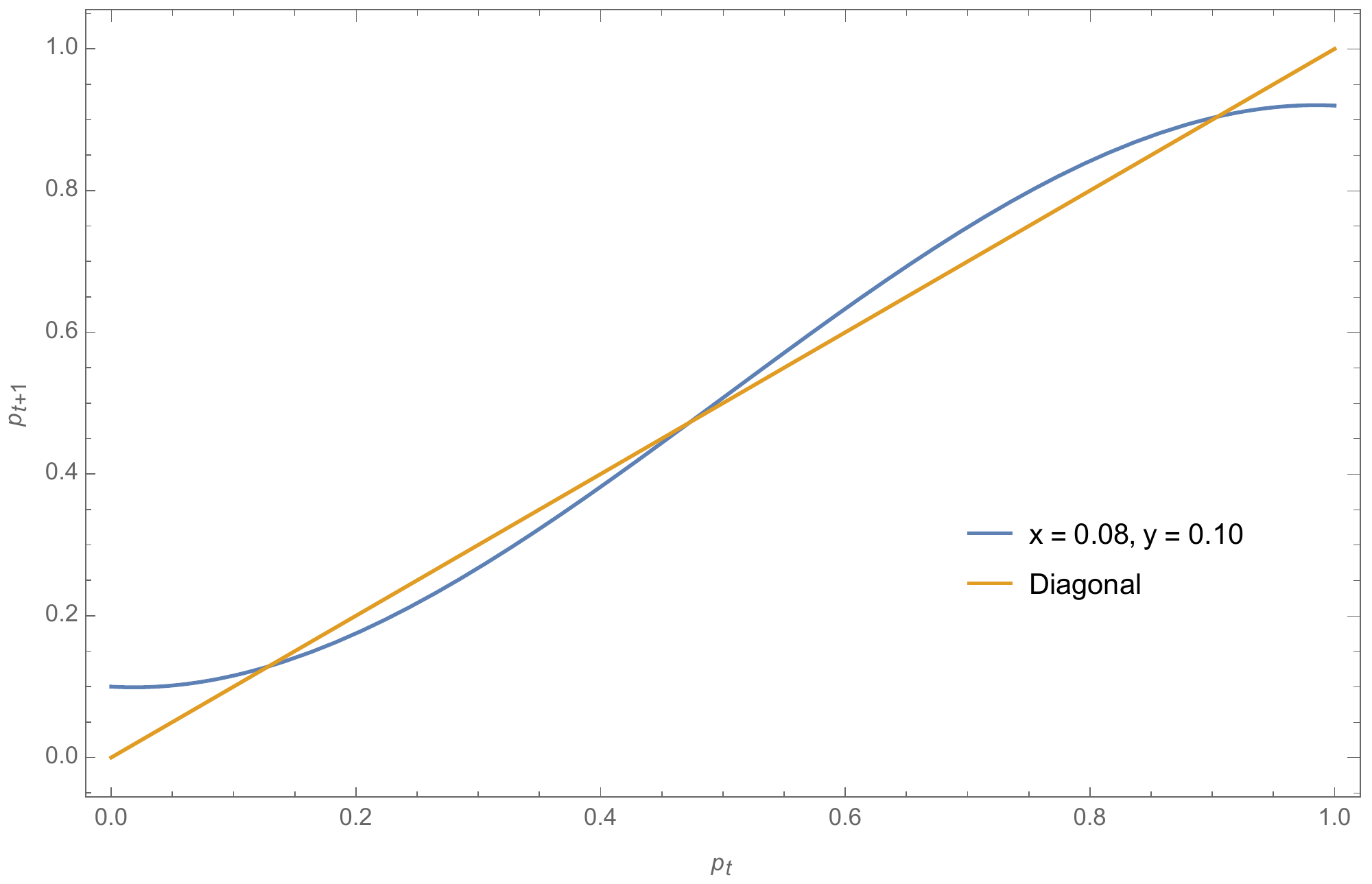}\ \ \ \ 
\includegraphics[width=.41\textwidth]{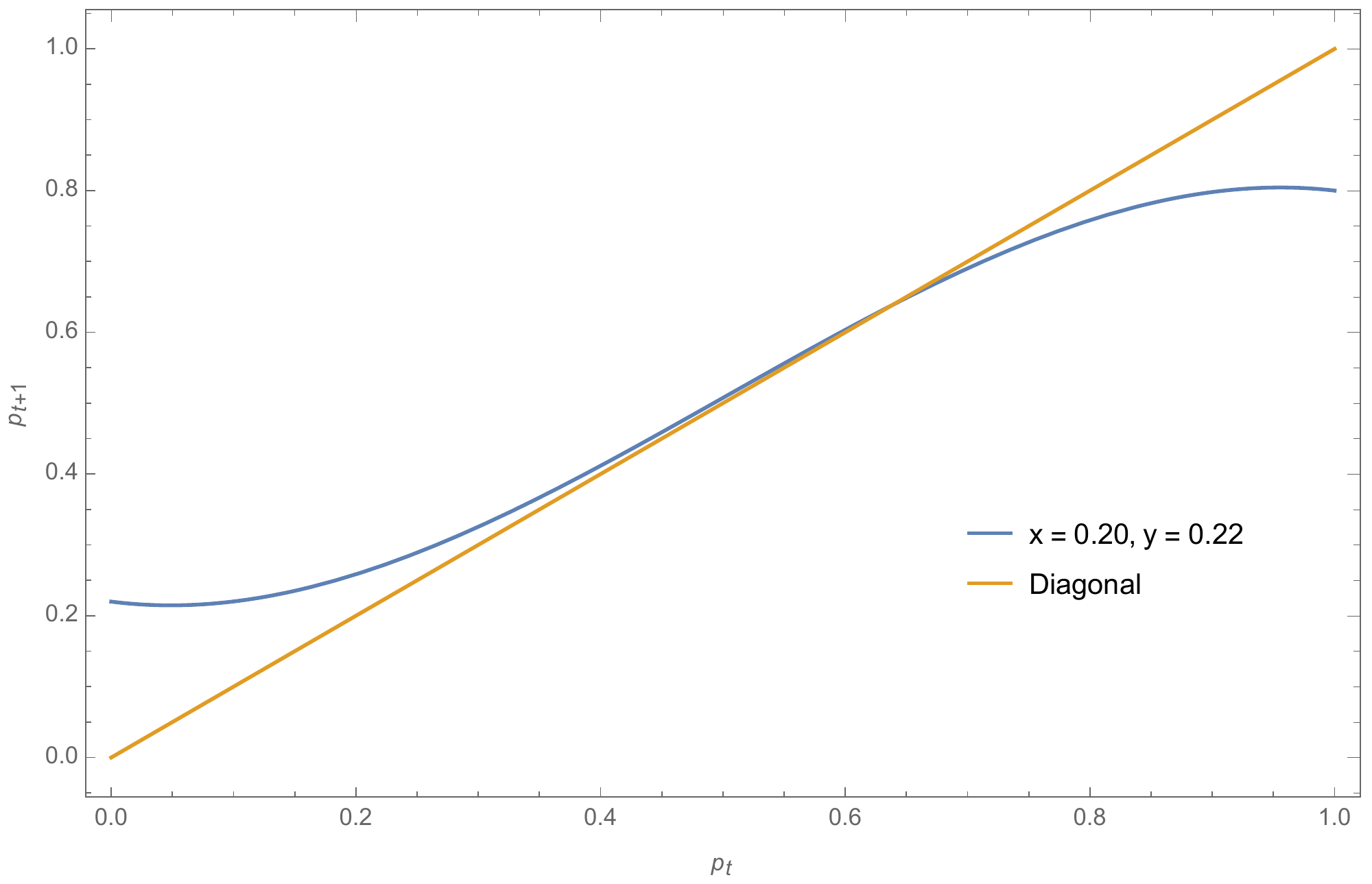}
\caption{Using Eq. (\ref{p15}) the expression $p_{t+1}$ as a function of $p_t$ is shown for $x=0.08, y=0.10$ (upper curve) and $x=0.20, y=0.22$ (lower curve).
The associated fixed points are respectively $p_A=0.13, p_t=0.47, p_B=0.90$ and $p_A=p_t=p_B=0.64$.} 
\label{f7f8}
\end{figure}

For $x=0.08$ and  $y=0.10$, Eq. (\ref{p15}) yields $p_A=0.13, p_t=0.47, p_B=0.90$. The coalescence phenomena observed with symmetrical contrarians is still valid but now it occurs at a value different from $\frac{1}{2}$ as seen from the case $x=0.20, y=0.22$ yielding $p_A=p_t=p_B=0.64$. These two cases are exhibited in Figure  (\ref{f7f8}).

\section{Discussion}

Asymmetric  contrarians were shown to generate four different schemes to account for the effect on the associated opinion dynamics. For update groups of size three, the various components to the update equations are listed in Table (\ref{ta5}). Combined together, the respective update equations are,
\begin{eqnarray}
P_{x,y}^1(p) & = & (1-x) \Big[ p^3  + 3 p^2(1-p)\Big] +  y  \Big[ 3 p(1-p)^2 + (1-p)^3\Big], \nonumber \\
P_{x,}^{ 2a}(p) & = & (1-x) \Big[ p^3  + 3 p^2(1-p)\Big] \Leftarrow  p>\frac{1}{2}, \nonumber\\
P_{y}^{2b}(p) & = &  p^3  + 3 p^2(1-p)+  y  \Big[ 3 p(1-p)^2 + (1-p)^3\Big]  \Leftarrow  p<\frac{1}{2}, \nonumber\\
P_{x,y}^3(p) & = & (1-x) p^3  +\Big[ 1- \frac{2}{3} x - \frac{1}{3} y) \Big] 3 p^2(1-p)  \nonumber\\
& + & \Big[ \frac{2}{3} y + \frac{1}{3} x\Big] 3 p(1-p)^2  +  y (1-p)^3\ , \nonumber\\
P_{x,y}^4(p) & = & (1-x) p^3  +  \Big[ 1- \frac{2}{3} x) \Big] 3 p^2(1-p) \nonumber\\
& + & \Big[ \frac{2}{3} y\Big] 3 p(1-p)^2 +y (1-p)^3.
\label{p-all}
\end{eqnarray}

\begin{table}
\centering
\begin{tabular}{|c|c|c|c|c|c|}
 \hline
 & Configuration & A A A  & A A B & A B B & B B B \nonumber\\
 \hline
 & Probability & $p^3$ & $3p^2(1-p)$  & $3 p(1-p)^2$ & $(1-p)^3$  \nonumber\\
 \hline
 \hline
1 &  post-update local majority & $(1-x)$ & $(1-x)$ & $y$ & $y$ \nonumber\\
 \hline
2a& post-update global majority $p>\frac{1}{2}$ & $ (1-x)$ & $(1-x) $ & $ 0$ & $ 0$ \nonumber\\
 \hline
2b& post-update global majority $p<\frac{1}{2}$ & $ 1 $ & $1 $ & $ y $ & $ y $ \nonumber\\
 \hline
 \hline
3 &In-groups local majority/minority  & $ (1-x) $ & $ (1-\frac{2}{3} x -\frac{1}{3} y)$ & $(\frac{2}{3} y + \frac{1}{3} x)$ & $y $ \nonumber\\
 \hline
4 &In-groups local majority  & $ (1-x) $ & $ (1-\frac{2}{3} x )$ & $\frac{2}{3} y $ & $y $ \nonumber\\
 \hline
 \end{tabular}
\caption{Rescaling factors for the probability of occurrence of configurations A A A, A A B, A B B, B B B for the various contrarian schemes.}
\label{ta5}
\end{table}

The common feature of four schemes is the breaking of the dynamics symmetry. In particular, the fifty-fifty attractor is shifted to a value different from fifty-fifty yielding a majority and a minority. Comparing schemes 1, 3 and 4 shows some minor quantitative differences while scheme 4 is rather different as expected since it is connected to the global majority. 

It is worth to notice that while $P_{x,x}^1(p) =P_{x,x}^3(p) \neq P_{x,x}^4(p)$ since indeed scheme 4 activates contrarian behavior only when an agent is at minority within an updating group before the update is completed.

\begin{figure}
\centering
\includegraphics[width=.46\textwidth]{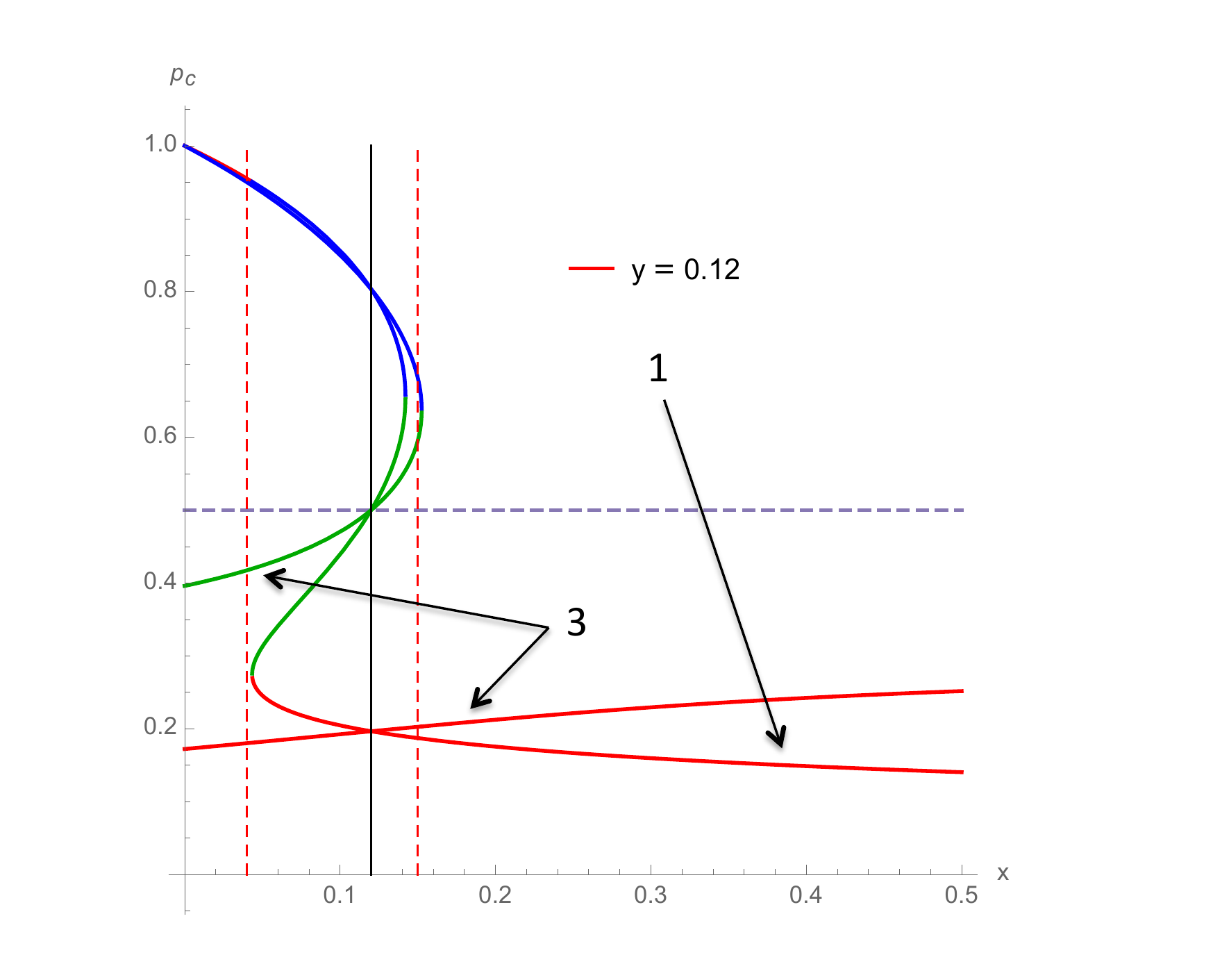}\ \ \ \ 
\includegraphics[width=.38\textwidth]{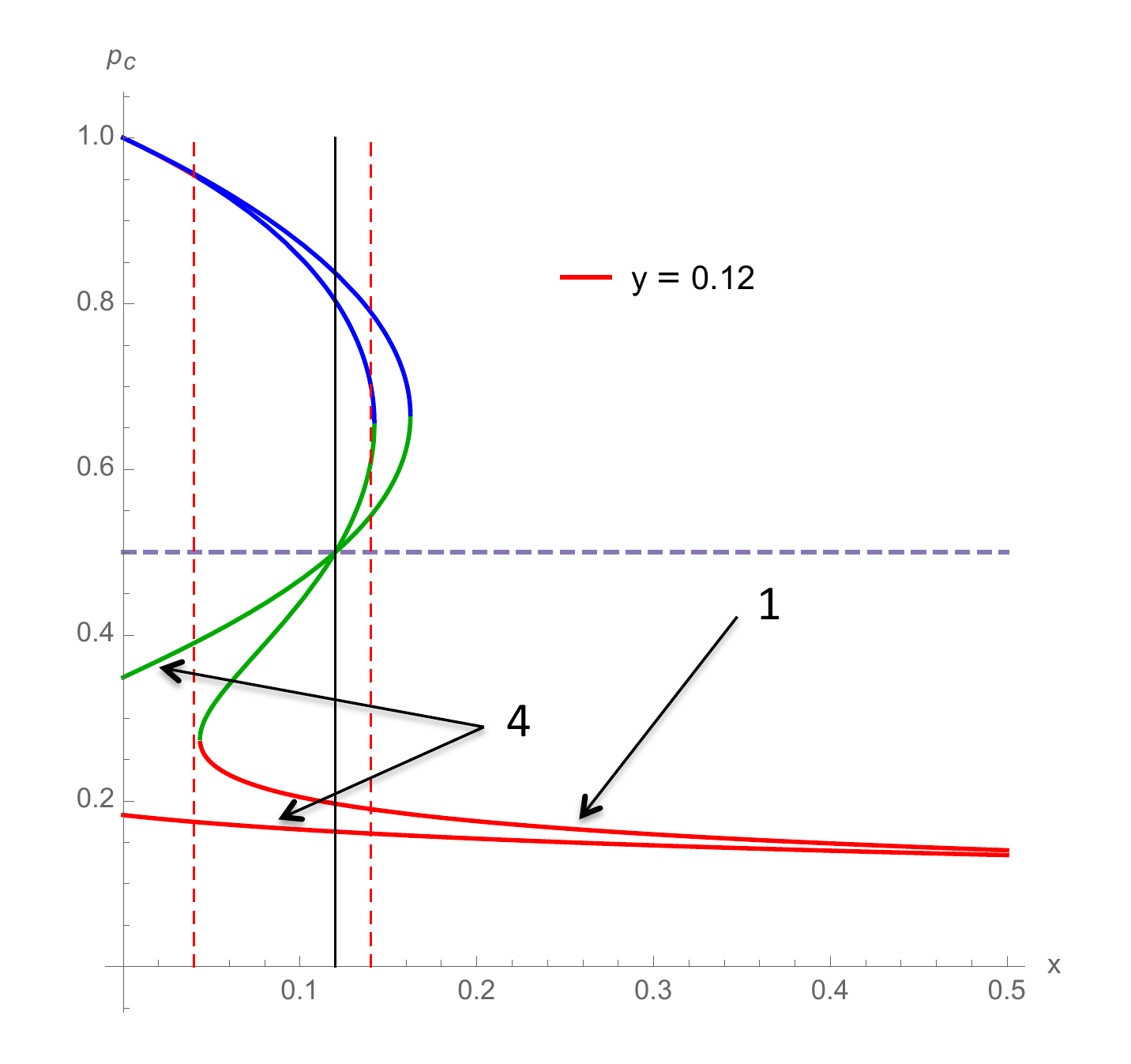}
\caption{Fixed points of scheme 1 and 3 (left) and schemes 1 and 4 (right) as a function of $x$ for $y=0.12$. The two dotted vertical lines show the area with a tipping point between two attractors. Outside the lines, the dynamics is driven by one single attractor.} 
\label{t1}
\end{figure}

\begin{figure}
\centering
\includegraphics[width=.39\textwidth]{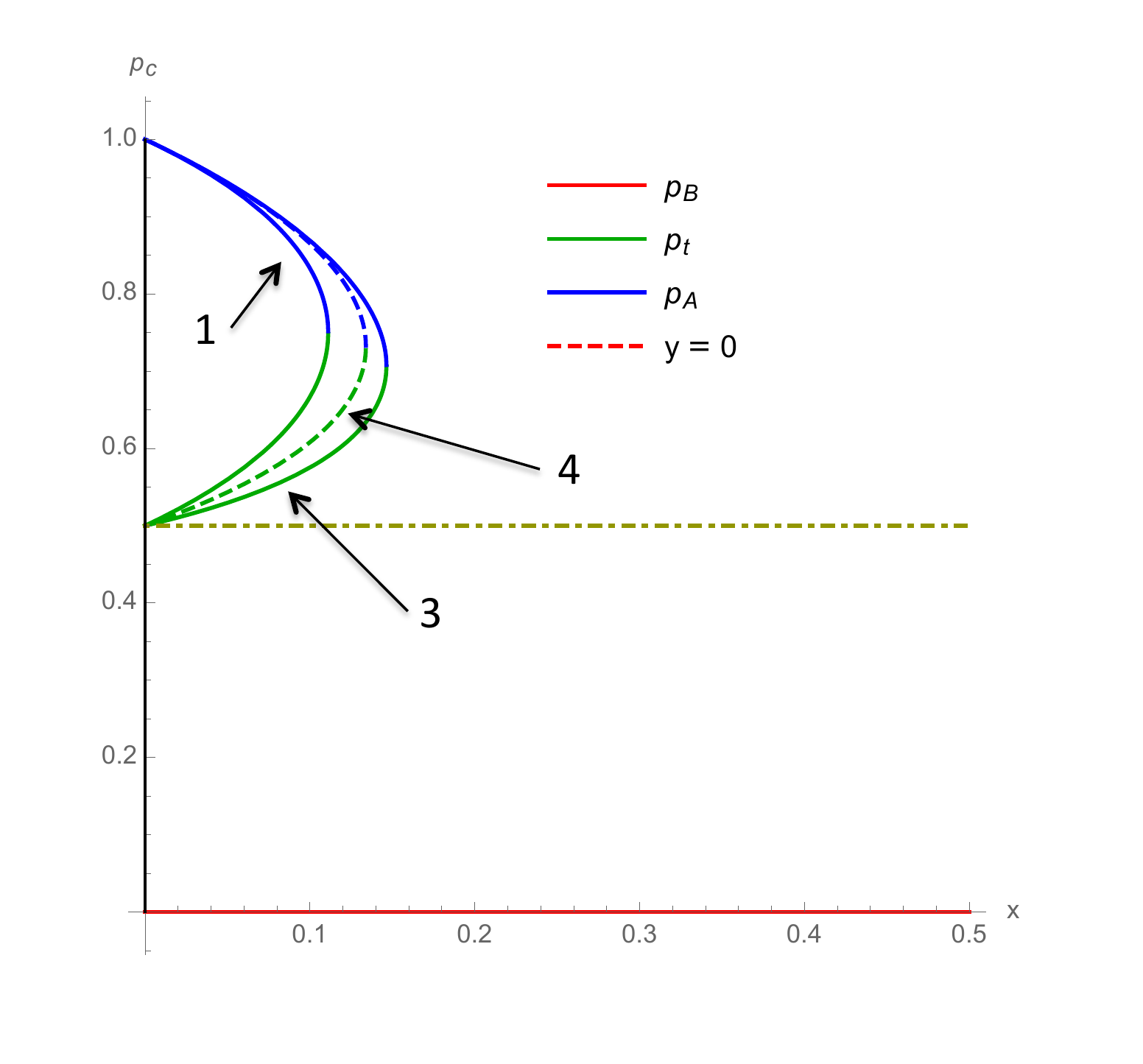}\ \ \ \ 
\includegraphics[width=.38\textwidth]{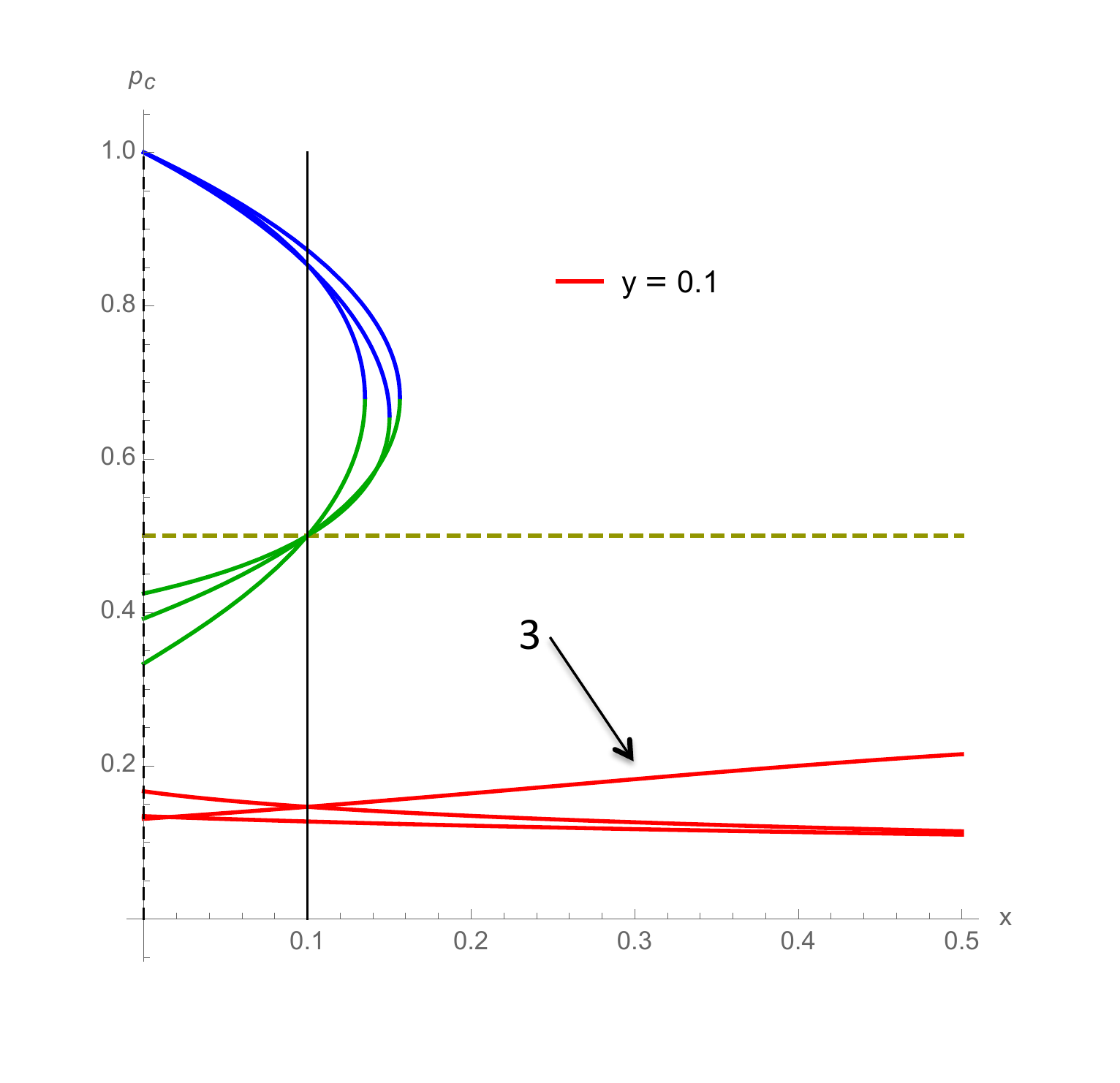}
\caption{Fixed points of schemes 1 and 3 and 4 as a function of $x$ for $y=0$ (left) and $y=0.10$ (right).} 
\label{t2}
\end{figure}

\begin{figure}
\centering
\includegraphics[width=.38\textwidth]{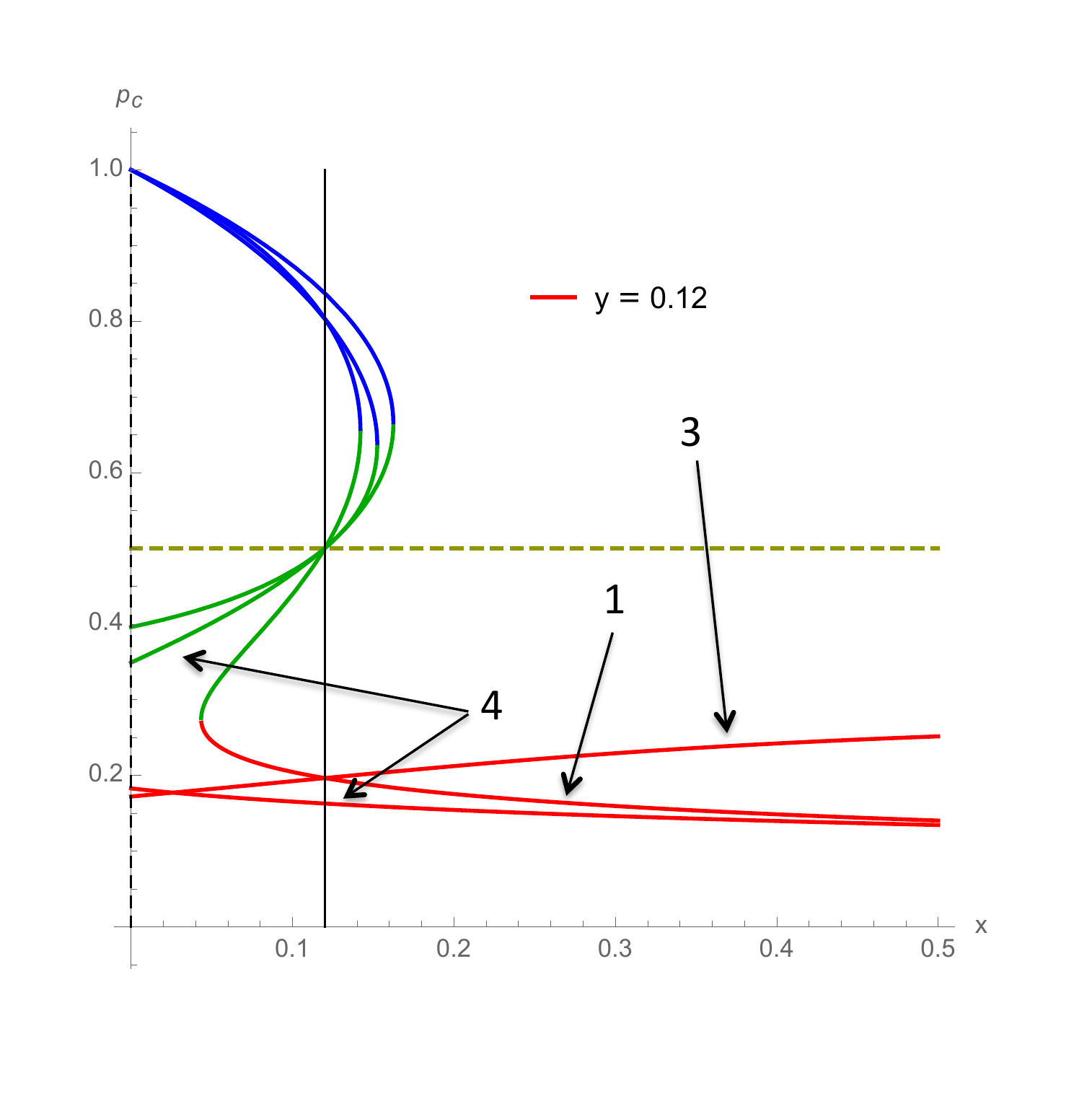}\ \ \ \ 
\includegraphics[width=.39\textwidth]{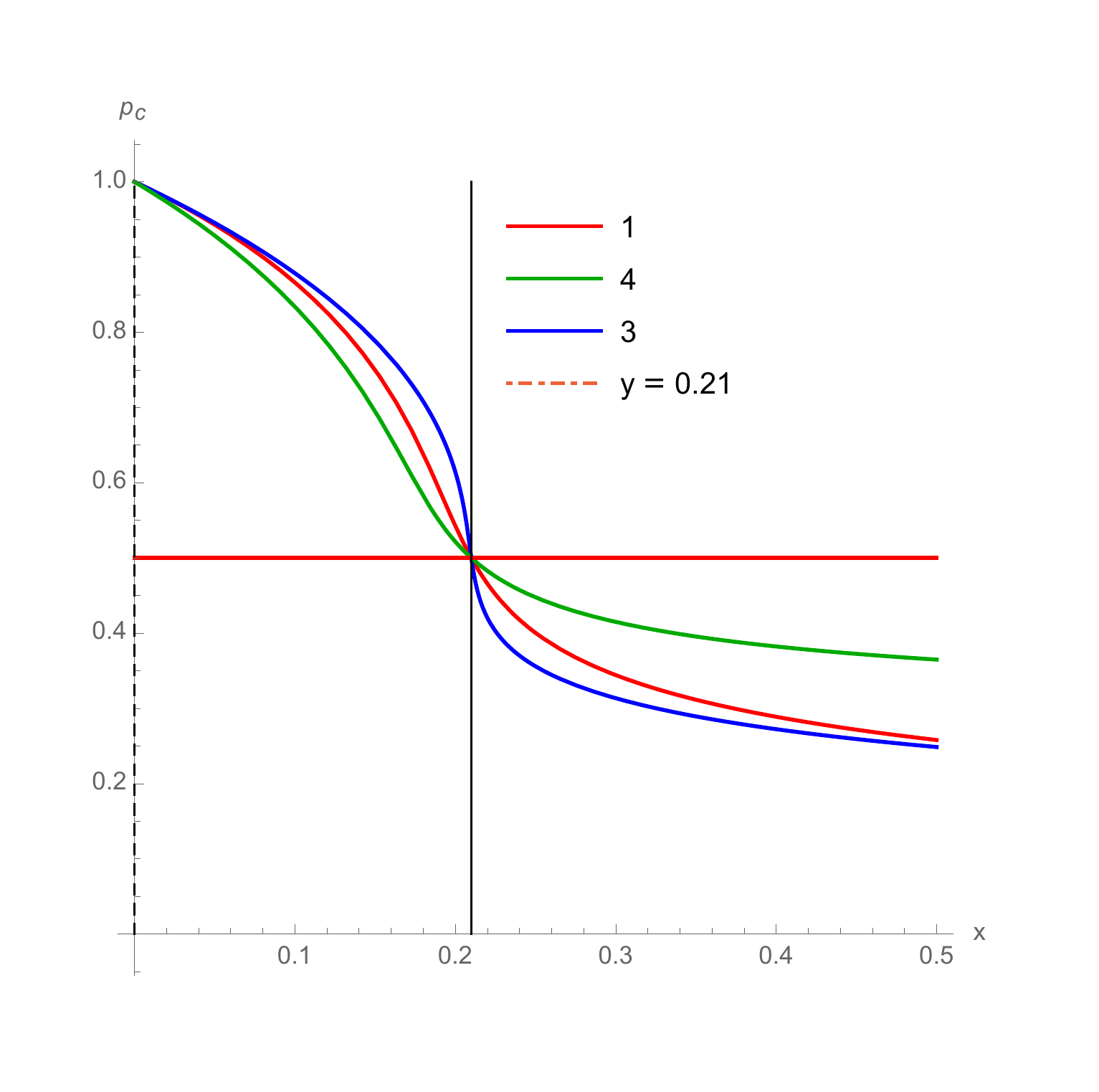}
\caption{Fixed points of schemes 1 and 3 and 4 as a function of $x$ for $y=0.12$ (left) and $y=0.21$ (right).} 
\label{t3}
\end{figure}

To illustrate the differences between the various schemes 1, 3 and 4, we show several cases of variations of the associated fixed points as a function of $x$ for a fixed value of $y$. Figure (\ref{t1}) compares schemes 1 and 3 and 1 and 4 for $y=0.12$. Figure (\ref{t2}) compares schemes 1, 3 and 4  for $y=0$ and $y=0.10$ while Figure (\ref{t3}) does the same for $y=0.12$ and $y=0.21$.

Left part of Figure (\ref{t1}) and right part of Figure (\ref{t2}) show that for scheme 3 the attractor $p_B$ increases slightly its value with increasing $x$, which looks paradoxical since contrarians are expected to decrease the overall support of their corresponding opinion.  Indeed, scheme 3 maintains a contrarian being in a minority position to keep on its opinion. It implies that when support for A (B) is low, many configurations have A (B) in minority and thus preserves A (B) opinion. More A (B) contrarians for low $p$ thus increases slightly the support for A (B).

\begin{figure}
\centering
\includegraphics[width=.41\textwidth]{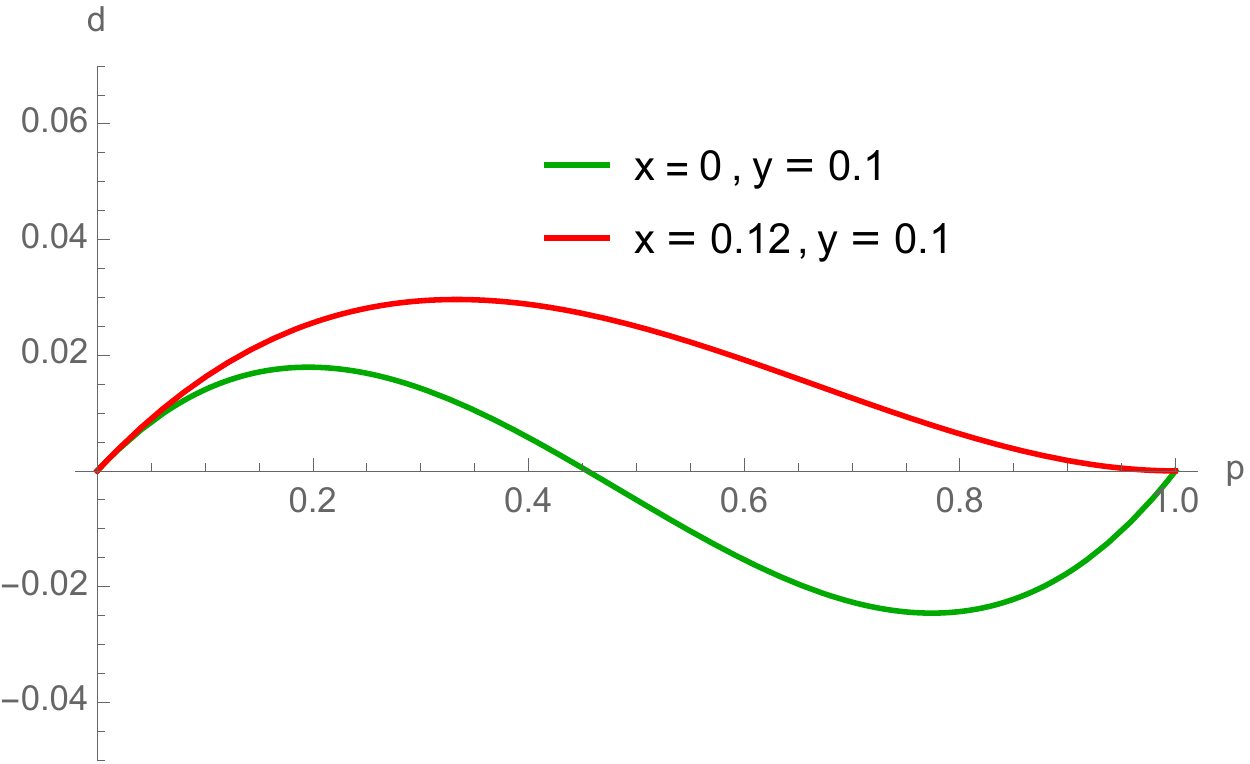}\ \ \ \ 
\includegraphics[width=.41\textwidth]{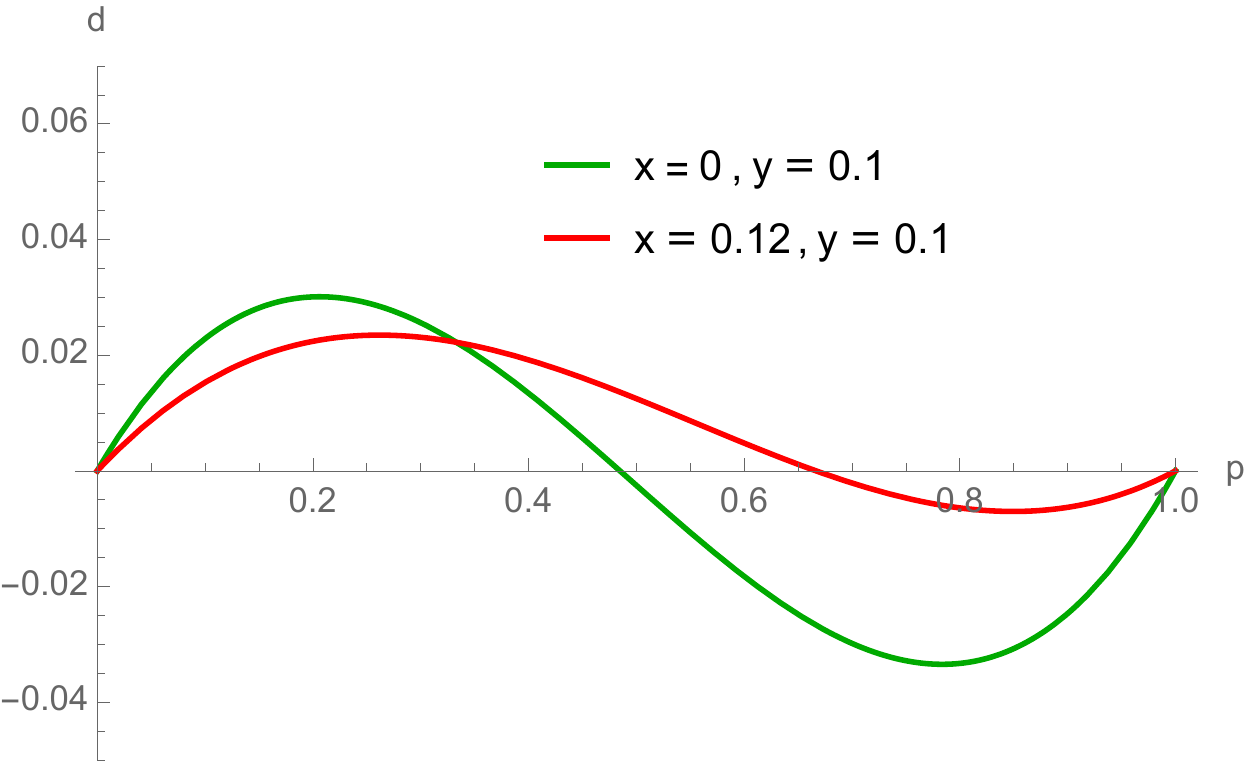}
\caption{Net balance between gain minus loss from contrarian behavior for scheme 4 and 3 when $x=0$ and $y=0.10$ (left) and $x=0.12$ and $y=0.10$ (right). More A contrarians for low values of $p$ increases the net gain for A.} 
\label{t4}
\end{figure}

Figure  (\ref{t4})  provides an illustration of this paradox exhibiting the balance between the loss and gain from contrarian behavior as a function of $p$, from configurations A A B and A B B for schemes 3 (right) and 4 (left). The cases $x=0, y=0.10$ and  $x=0,12, y=0.10$ are compared. It is seen that increasing $x$ reduces the net balance at low $p$ for scheme 4 as expected. However, the opposite occurs for scheme 3.

\begin{figure}
\centering
\includegraphics[width=.41\textwidth]{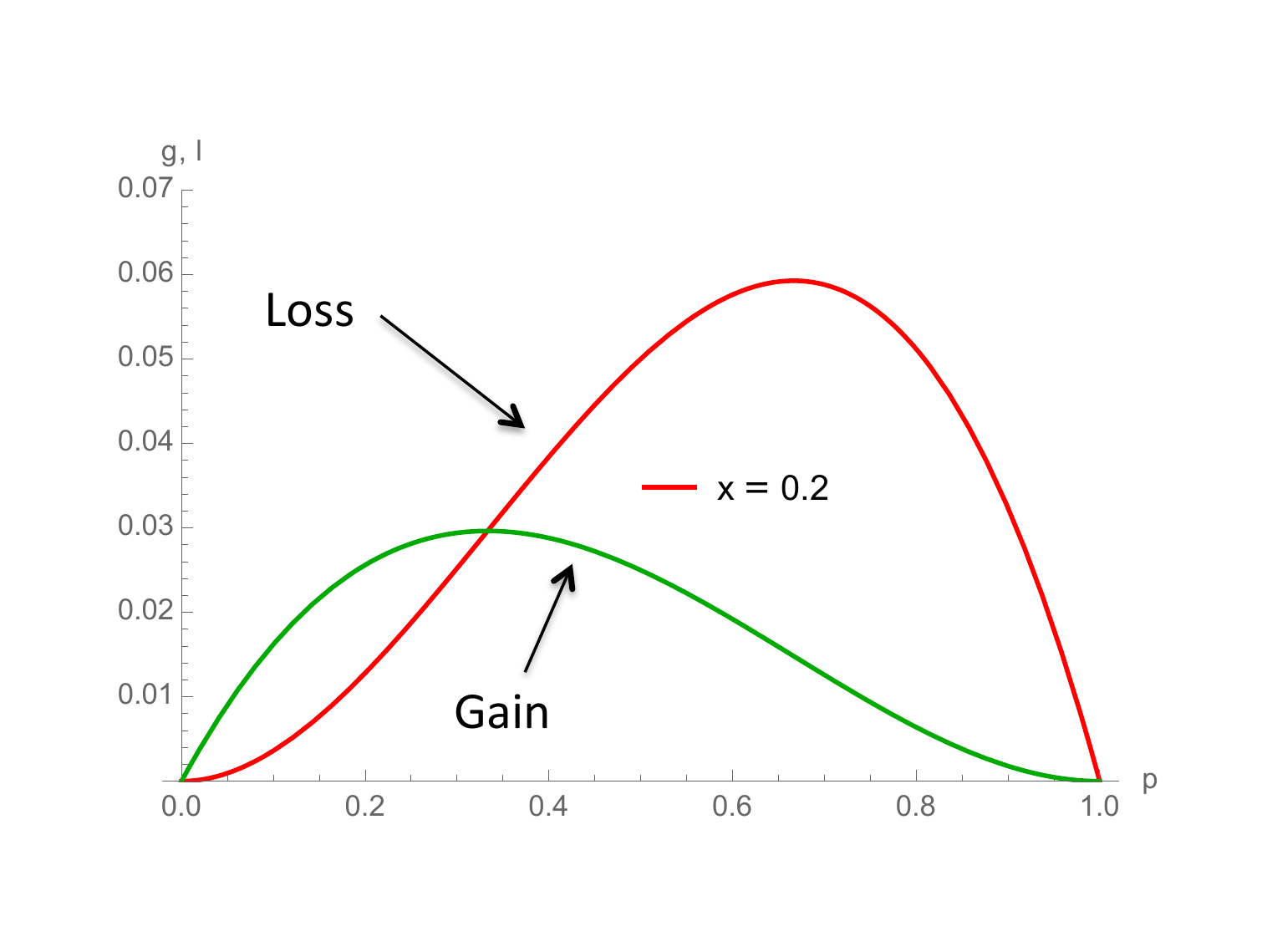}
\caption{Amplitude of the negative contribution $- \frac{2}{3} x  3 p^2(1-p)$ (Loss) from A contrarians (proportion $x$) and the amplitude of the positive contribution $\frac{1}{3} x  3 p (1-p)^2$ (Gain) for $x=0.20$. The positive contribution is seen to be bigger than the negative one for $p<\frac{1}{3}$.} 
\label{t5}
\end{figure}

Indeed, looking at the 4 schemes associated with the update equations listed in Eq. (\ref{p-all}) , it is seen that only $P_{x,y}^3(p) $ has a positive $x$ contribution from the minority configurations A B B, which explains the counterintuitive finding that at low values of $p$, increasing $x$ increases a bit the A support. Figures (\ref{t4},\ref{t5})) shows both the amplitude of the negative contribution $- \frac{2}{3} x  3 p^2(1-p)$ from A contrarians (proportion $x$) and the amplitude of the positive contribution $\frac{1}{3} x  3 p (1-p)^2$ for $x=0.20$. The positive contribution is seen to be bigger than the negative one for $p<\frac{1}{3}$.

\section{Conclusion}

We have introduces asymmetry in contrarian behavior with respect to two competing opinions A and B to account for inhomogeneity in the distribution of psychological traits within a given population in connexion to the ability to differ from the group. It hints at contrarian behavior being connected to the underlying cultural frame of  social groups. Accordingly, symmetric contrarians are expected to prevail within homogeneous populations.

Asymmetry was shown to generate several different schemes for the contrarian activation. A rich variety of changes has been found about the values of the attractors and tipping point \cite{metrics}.
 
However, the differences by themselves are not of large amplitude establishing the robustness of contrarian behavior. Nevertheless, it is worth to emphasize that as soon as asymmetry is turned on, the salient feature of having a single attractor located precisely at fifty-fifty is waived off. However, now a slightly lower proportion of contrarian on its own side than on the other competing side is the key to eventually win a public debate, whin in turn can transform into an election victory. This result allows to envision novel wining strategies focusing on the lowering of the proportion of contrarians on its own side.

\end{document}